\documentclass[10pt, twocolumn]{IEEEtran} 
\usepackage[T1]{fontenc}
\usepackage{color}
\usepackage{mathrsfs}
\usepackage{amsmath}
\usepackage{amsthm} 
\usepackage{amssymb}
\usepackage{graphicx}
\usepackage{cite}
\usepackage{makecell}
\usepackage{algorithm}
\usepackage{algorithmic}
\usepackage{fancyhdr}
\usepackage{diagbox}
\usepackage{bm}
\usepackage[unicode=true,
 bookmarks=true,bookmarksnumbered=true,bookmarksopen=true,bookmarksopenlevel=1,
 breaklinks=false,pdfborder={0 0 0},pdfborderstyle={},backref=false,colorlinks=false]
 {hyperref}
\hypersetup{
 pdftitle={Your Title},
 pdfauthor={Your Name},
 pdfpagelayout=OneColumn, pdfnewwindow=true, pdfstartview=XYZ, plainpages=false}

\usepackage{setspace}
\usepackage{multirow} 
\usepackage{xcolor}

\allowdisplaybreaks[4]

\usepackage[caption=false,font=footnotesize]{subfig}

\IEEEoverridecommandlockouts

\usepackage{lettrine}
\usepackage{geometry}
\renewcommand{\headrulewidth}{0pt} 
\renewcommand{\arraystretch}{1.2}

\begin{document}

\title{\textcolor{black}{
    Trusted Routing for Blockchain-Empowered UAV Networks via Multi-Agent Deep Reinforcement Learning
    }}   
\author{
    \IEEEauthorblockN{
    Ziye Jia, \IEEEmembership{Member, IEEE}, 
    Sijie He, 
    Qiuming Zhu, \IEEEmembership{Senior Member, IEEE}, 
    Wei Wang, \IEEEmembership{Member, IEEE}, \\
    Qihui Wu, \IEEEmembership{Fellow, IEEE}, 
    and Zhu Han, \IEEEmembership{Fellow, IEEE}}
    \thanks{Ziye Jia is with the College of Electronic and Information Engineering, 
    Nanjing University of Aeronautics and Astronautics, Nanjing 211106, China,
    and also with the State Key Laboratory of Integrated Service Networks, 
    Xidian University, Xi'an Shaanxi 710071, China (e-mail: jiaziye@nuaa.edu.cn).}
    \thanks{Sijie He, Qiuming Zhu, Wei Wang, and Qihui Wu are with the College of Electronic 
    and Information Engineering, Nanjing University of Aeronautics and Astronautics, 
    Nanjing 211106, China (e-mail: sx2304109@nuaa.edu.cn; zhuqiuming@nuaa.edu.cn;
    {wei\_wang@nuaa.edu.cn}; wuqihui@nuaa.edu.cn).}
    \thanks{Zhu Han is with the University of Houston, Houston, 
    TX 77004 USA, and also with the Department of Computer Science and Engineering, 
    Kyung Hee University, Seoul 446-701, South Korea (e-mail: hanzhu22@gmail.com).}
    }

\maketitle
\pagestyle{fancy}
\fancyhf{}
\fancyhead[R]{\fontsize{7}{9}\selectfont \thepage}
\renewcommand{\headrulewidth}{0pt} 
\renewcommand{\footrulewidth}{0pt}

\begin{abstract}
    Due to the high flexibility and versatility, unmanned aerial vehicles (UAVs)
     are leveraged in various fields including surveillance and disaster rescue. 
    However, in UAV networks, routing is vulnerable to malicious damage due to 
    distributed topologies and high dynamics. Hence, ensuring the routing security 
    of UAV networks is challenging. In this paper, we characterize the routing 
    process in a time-varying UAV network with malicious nodes. Specifically, 
    we formulate the routing problem to minimize the total delay, which is an 
    integer linear programming and intractable to solve.     
    \textcolor{black}{
        Then, to tackle the network security issue, a blockchain-based trust 
        management mechanism (BTMM) is designed to dynamically evaluate trust values  
        and identify low-trust UAVs.
        To improve traditional practical Byzantine fault tolerance algorithms 
        in the blockchain, we propose a consensus UAV update mechanism.
        Besides, considering the local observability, 
        the routing problem is reformulated into a decentralized partially observable Markov 
        decision process.  
        Further, a multi-agent double deep Q-network based routing algorithm is designed to minimize the total delay.}
    Finally, simulations are conducted with attacked UAVs and numerical results 
    show that the delay of the proposed mechanism decreases by 13.39$\%$, 12.74$\%$, 
    and 16.6$\%$ than multi-agent proximal policy optimal algorithms, multi-agent 
    deep Q-network algorithms, and methods without BTMM, respectively.

\end{abstract}

\newcommand{\CLASSINPUTtoptextmargin}{0.8in}

\newcommand{\CLASSINPUTbottomtextmargin}{1in}

\section{Introduction}

\lettrine[lines=2]{A}{s} a key component of the space-air-ground integrated 
network in the six generation communication techniques, 
unmanned aerial vehicles (UAVs) are rapidly developed and widely applied 
to multiple tasks, 
such as disaster rescue and real-time monitoring{\cite{10418158, JiaHiera}}. 
In these applications, UAVs can serve as aerial base stations for 
data collection and transmission, providing low-cost, flexible,
and versatile services {\cite{10430396}}.
Besides, UAV networking can cooperatively accomplish complex 
tasks, in which routing is a significant issue for optimizing the 
data transmission {\cite{Routing_UAV_Survey, 10574195, LU_Zhou}}. 
Moreover, since UAV networks are characterized by the complex application
environment, high dynamic and distributed topology, they are vulnerable 
to malicious attacks {\cite{10430396}}.
When there exist malicious UAVs, 
the hostile activities can lead to the degradation of routing performances. 
Therefore, it is necessary to evaluate the behaviors and manage the malicious UAVs.
There exist a couple of works focusing on building the node trust table by  
a single-node based centralized manager, which is not resilient to fault 
tolerance and can be easily tampered with {\cite{Trust_survey9264256}}.
Hence, to improve the routing security, the blockchain technique has been 
introduced to build decentralized trusted mechanisms  {\cite{Blockchain9120287}}. 
In particular, the blockchain is a distributed ledger, consisting of a chain 
of blocks, in which transaction details and records are stored and 
hard to be tempered with {\cite{blockchai_security}}.


Meanwhile, delay is a key quality of service for routing in the UAV network,
since low delay can significantly improve the timeliness and reliability for 
various emergency applications, such as surveillance information transmission {\cite{ Wang_security}}.
Besides, the energy and storage capacity of UAV networks 
are constrained. Hence, it is necessary to collaborate the resources of 
UAV networks to provide routing services with low delay.
Additionally, due to the distributed topology, high mobility,
and low-trust of UAV networks, the availability of 
communication links is susceptible, leading to the reduction of routing performances. 
Therefore, it is crucial to design a dynamic routing 
algorithm to maintain reliable and robust communications in UAV networks. 
Most traditional routing algorithms are designed for static environment or fixed rules, 
such as the Bellman Ford algorithm and Dijkstra algorithm, which cannot be directly applied
 to UAV networks {\cite{Bellman-Ford-Dijkstra,10184319}}.
Besides, it is challenging to obtain the global 
information of the decentralized UAV network {\cite{JiaCooper}}.
Hence, it is significant to design a dynamic and adaptive algorithm 
to enable the efficient and safe routing in UAV networks.

To deal with the above challenges, in this paper, due to the 
dynamic and distributed topology, as well as the complex 
environment of UAV networks, the routing process is depicted by 
a time-varying model with malicious UAVs.
 In light of the constructed UAV network,
the routing problem is formulated to minimize the total end-to-end delay, 
which is an integer linear programming (ILP) and intractable to solve. 
Besides, to ensure the security of routing among UAVs, a node trust management 
mechanism based on the blockchain technique is proposed, 
in which each UAV obtains a trust value with respect to the delivery rate 
and the correctness of transmission paths. When the value is lower 
than the given safety threshold, the UAV is isolated from the network.
Besides, it is challenging to obtain the global 
information due to the decentralized UAV network.
Hence, we reformulate the routing problem into a decentralized partially 
observable Markov decision process (Dec-POMDP) and further design a multi-agent double deep
Q-network (MADDQN)-based approach. 
\textcolor{black}{ 
Specially, Dec-POMDP is a mathematical framework from the variant of the MDP, 
for modeling the decentralized multi-agent and partial observability of the environment.
In this work, decentralized UAVs act as agents that independently observe local states 
and jointly optimize routing decisions {\cite{Dec-POMDP}}.}
Finally, numerical simulations are conducted to verify the performance of the proposed algorithms.

In short, the major contributions of the proposed work are summarized as
follows:
\begin{itemize}

    \item \textcolor{black}{ We characterize the routing process via a time-varying
    UAV network with malicious nodes. Besides, we formulate the routing 
    problem to minimize the total end-to-end delay, which is an ILP and
    tricky to solve by traditional optimization schemes.}
    
    \item \textcolor{black}{ To deal with the issue of network security, the blockchain-based trust management 
    mechanism (BTMM) is designed, which records the malicious behaviors and 
    trust values of UAVs. Further, to promote the routing security in UAV
    networks, we design the consensus node update mechanism to improve  the
    practical Byzantine fault tolerance (PBFT) algorithms. }
    
    \item \textcolor{black}{ 
    Since it is challenging to obtain the global information due to the decentralized
    UAV networks, we reformulate the routing problem into a Dec-POMDP. Furthermore, an 
    MADDQN-based routing algorithm is proposed to learn the dynamically changing
    network topology and make decisions to minimize the end-to-end routing delay.}

    \item \textcolor{black}{Extensive simulations are conducted to evaluate the proposed BTMM-MADDQN 
    algorithms with attacked UAVs. Numerical results show that the proposed
    algorithm reduces the delay compared to the BTMM-multi-agent proximal policy 
    optimal (BTMM-MAPPO) algorithm, BTMM-multi-agent deep Q-network (BTMM-MADQN) 
    algorithm, and MADDQN algorithm without BTMM when there exist malicious UAVs.}
\end{itemize}

    The rest of this paper is organized as follows. 
    In Section {\ref{Related-works}}, we introduce the related works. 
    The system model and optimization problem are illustrated in Section {\ref{System-model}}. 
    Then, we design the algorithms in Section {\ref{sec:Reinforcement-learning}}. 
    Simulations are conducted and the performance analyses are shown in Section {\ref{sec:Simulation Results}}.
    Finally, the conclusions are drawn in Section {\ref{sec:Conclusions}}.
 
    \begin{figure*}[t]
        \vspace{-0.1cm}
        \centering
        \includegraphics[width=0.8\linewidth]{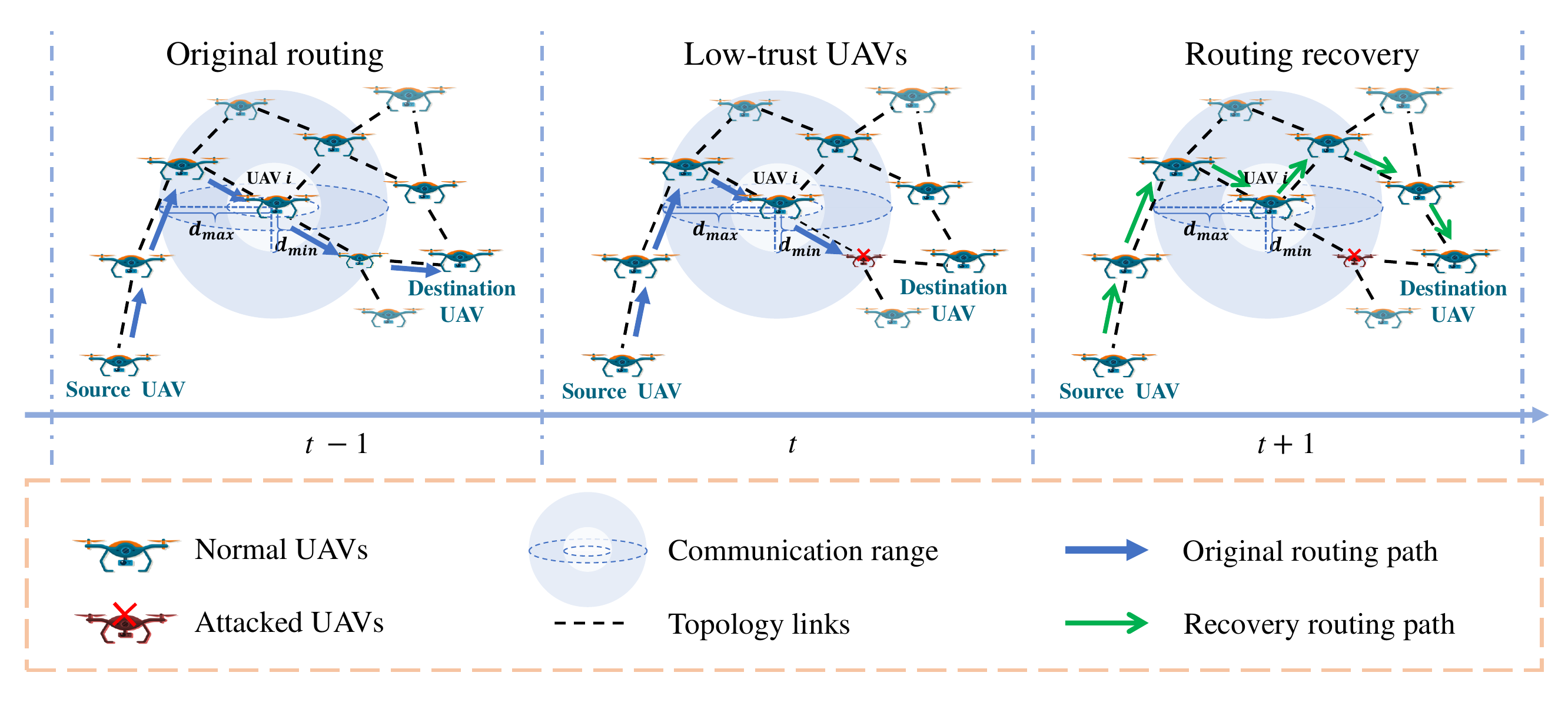}
        \textcolor{black}{\caption{\label{fig:Network_of_UAV}Routing recovery in a time-varying UAV network. 
        }}
        \vspace{-0.1cm}
    \end{figure*}
\section{Related Works\label{Related-works}}
There exist a couple of works related to the trust management. 
For instance,  the authors in {\cite{Trust_9619480}} propose trust attestation 
mechanisms based on a centralized authority to evaluate the trust of a network
slice. To promote the security and efficiency for edge devices, 
{\cite{Trust_10422726}} presents a dynamic trust management model based on the feedback, 
in which a third-party trust agent is set up to cooperatively manage the trust and
a cloud center is assumed to be reliable.
Authors in {\cite{ETERS}} evaluate the trust via a multi-trust method, 
in which the base station is considered as a central command authority 
that cannot be compromised by an attacker. 
In {\cite{LightTrust_9434372}}, a lightweight trust management is presented to overcome the 
security issue, which utilizes a centralized trust agent to maintain 
a database that stores the trust degree. As above, the centralized trust 
management approaches are not applicable to distributed UAV networks.
Although these works can enhance the network security by managing the node trust, 
most of them rely on a centralized manager, which is susceptible
to single point of failure and being tampered with.

Therefore, with the decentralized and tamper-proof characteristics,  
the blockchain technique attracts more attentions to secure networks.
In {\cite{DRRS-BC}}, a blockchain-based decentralized registration framework 
is designed to solve the problem of misconfiguration and malicious attacks. 
The authors propose a trusted network framework for flying ad-hoc networks, 
utilizing the blockchain technique to realize the distributed peer-to-peer security {\cite{Blockchain_Empowered}}.
The authors in {\cite{Trusted_Blockchain}} introduce a trusted self-organizing
framework of UAV networks based on the blockchain to enhance the internal security.
In {\cite{Blockchain_Date_Sharing}}, the authors present an efficient and secure
data sharing approach for Internet of things via adopting the blockchain. 
{\cite{Access_Blockchain}} presents a private blockchain-enabled 
fine-grained access control mechanism for the software defined network environment. 
The authors in {\cite{IIoT_Blockchain}} propose a blockchain-based access control
framework for industrial Internet of things. The above works illustrate that the 
blockchain technique enables to improve the network safety, and it is possible 
to employ the blockchain to ensure routing security in UAV networks.
As a significant issue, routing algorithms have been  
widely studied. 
For instance, the topology-based routing methods pose heavy communication overloads 
 {\cite{10077737}}. 
The greedy geographic routing methods are susceptible to falling into 
the local optimal solution {\cite{Routing7323861}}.
The bio-inspired routing approaches struggle with parameter sensitivity and 
slow convergence {\cite{bio_routing}}.
Additionally, the mentioned algorithms are not suitable for 
distributed and highly dynamic UAV networks, where nodes can only obtain  
the localized information and topologies change constantly. 
Hence, it is significant to design a routing algorithm,  
effectively adapting to dynamics and local information.
With the advance in reinforcement learning (RL) techniques, 
there exist a lot of works related to RL-based routing methods. 
For example, authors in {\cite{Q-learning-based1}} utilize 
Q-learning algorithms to learn routing decisions, 
and the method is further improved in 
{\cite{Q-learning-based3, peng2022deep}}.
The authors in {\cite{DQN-2}} propose a MADQN-based algorithm to overcome 
the interference and link instability caused by line-of-sight communication links. 
{\cite{Q-learning-based3}} proposes a distributed multi-agent deep RL (DRL)-based  routing
solution to overcome link layer jamming attacks. 
Authors in {\cite{9410247}} investigate the MAPPO and 
multi-agent meta-PPO to optimize the network performances under both 
fixed and time-varying demands, respectively.
As above, RL-based mechanisms can solve 
routing problems effectively. Nevertheless, most of the above works lack considerations 
of the security in a high dynamic and trustless UAV network with malicious nodes. 

Hence, in this work, we consider utilizing the blockchain for improving 
the security of UAV networks with malicious nodes, and employ DRL 
techniques to deal with the trusted routing issues.  

\section{System Model and Problem Formulation\label{System-model}}
\subsection{Network Model}
As show in Fig. \ref{fig:Network_of_UAV}, we depict the routing process 
in a time-varying UAV network, in which $N$ nodes are distributed randomly, and 
each UAV holds a unique identity. 
Further, the UAV network is modeled as a weighted undirected 
graph $\mathcal{G}(\mathcal{U} ,\mathcal{E})$, 
where $\mathcal{U}$
denotes the set of all UAVs, and $\mathcal{U}_\mathtt{s}$, $\mathcal{U}_\mathtt{r}$, and 
$\mathcal{U}_\mathtt{d}$ indicate the sets of source, relay, and destination UAVs, respectively.
The values of $|\mathcal{U}_\mathtt{s}|$ and $|\mathcal{U}_\mathtt{d}|$ is equal to the number of demand set $R$.
$\mathcal{E}$ indicates the set of  communication link status among UAVs.
We utilize $e(i,j)\in\mathcal{E}$ to represent whether there exists a communicable link between 
UAVs $i$ and $j$ ($i\neq j \text{, } \forall i,j\in\mathcal{U}$). 
Specifically, $e(i,j)=1$ indicates the link is effective, 
and $e(i,j)=0$ denotes there exist no direct links between UAVs $i$ and $j$. 

In particular, the time period is divided into $T$ 
time slots, and the length of each time slot is $\tau$. 
With regard to demand $r\in R$, it is denoted as a tuple
\{$\mathtt{s},\mathtt{d},\mathcal{L}^r$\}, 
where $\mathtt{s}$, $\mathtt{d}$, and $\mathcal{L}^r$ 
indicate the source UAV, destination UAV, and data size of demand $r$, respectively. 
Moreover, $\mathtt{P}_{\mathtt{s} {\mathtt{d}}}^r={(e(\mathtt{s},i),\cdots,e(j,\mathtt{d}))}$ 
indicates a complete routing path for transmitting demand $r$ from 
source UAV $\mathtt{s} \in\mathcal{U}_\mathtt{s}$ to 
corresponding destination UAV $\mathtt{d} \in\mathcal{U}_\mathtt{d}$.
\textcolor{black}{In the UAV network model, for a specific demand, any two UAVs can serve as the source and destination nodes, 
while the remaining UAVs act as relay nodes to forward data.
It is noted that a relay node can have its own original demand to transmit and thus become a source node. 
It indicates that the roles of source, relay, and destination nodes are determined based on 
the specific data transmission requirements.
}

\textcolor{black}{
The location of UAV $i \in\mathcal{U}$ at time slot $t$ is indicated by 
${\varTheta}_{i(t)}={(x_{i(t)},y_{i(t)},z_{i(t)})}$ in three-dimensional
Cartesian coordinates.
The velocity in $x$, $y$, and $z$ directions of UAV $i$ is 
represented by ${{{\bm{\nu}}}}_{i(t)}={(\nu^{x}_{i(t)},\nu^{y}_{i(t)},\nu^{z}_{i(t)})}$, 
and the velocities of all UAVs are denoted as
 $\bm{V}(t)=\{\bm{\nu}_{1(t)},\bm{\nu}_{2(t)},\cdots,\bm{\nu}_{N(t)}\}$.  
Specifically, at the beginning of each time slot $t$, the UAV is able to instantaneously 
        adjust its flight direction and speed, 
        and subsequently fixes the flight direction and maintains a constant speed. 
       From a kinematic view, $\Theta_{i(t)}$ 
       is related to $\boldsymbol{\nu}_{i(t)}$ of UAV $i$.
       Specially, if UAV $i$ has a velocity of $\boldsymbol{\nu}_{i(t)}$ at the beginning of time slot $t$
       and maintains a uniform speed within the time slot, 
       its location change can be approximated as $\Delta\Theta_{i(t)} \approx \boldsymbol{\nu}_ {i(t)} \tau$.
       Therefore, at the next time slot,  the location $\Theta_{i(t + 1)}$ can 
       be computed as $\Theta_{i({t + 1})} = \Theta_{i(t)} + \boldsymbol{\nu}_{i(t)} \tau$.} 

    \textcolor{black}{    
        For data transmission, the location of the UAV determines the establishment 
        and maintenance of communication links.
        At time slot $t$, only when the Euclidean distance $d_{i(t) j(t)}$ between UAVs $i$ and $j$ 
        is within the communication range, a communication link can potentially be established for data transmission. Here,
                \begin{equation}{\label{distance}}
                    \begin{aligned}
                        &d_{i(t) j(t)}\!= \left\lVert \Theta_{i(t)}\!-\!\Theta_{j(t)}  \right\rVert _2 = \\
                        &\!\sqrt{(x_{i(t)} \!-\! x_{j(t)})^2\!+\!(y_{i(t)}\!-\!y_{j(t)})^2\!+\!(z_{i(t)}\!-\!z_{j(t)})^2},  
                    \end{aligned}
                \end{equation}
         and  the distance between UAV $i$ and possible connected UAV $k \in K_{i(t)}$ should satisfy 
        \begin{equation}{\label{distance_max}}
            d_{i(t) k(t)}\leqslant d_{i,max}, \forall k\in K_{i(t)}, i\in \mathcal{U}, t \in T,
        \end{equation}
        in which $d_{i,max}$ denotes the maximum communication distance of UAV $i$. 
        $K_{i(t)}$ indicates the set of UAVs that satisfies the constraint in (\ref{distance_max}) for UAV $i$ at time slot $t$.}
        \textcolor{black}{ 
        Additionally, $d_{i(t) j(t)}, \forall  i,j\in\mathcal{U}$ should satisfy
        \begin{equation}{\label{distance_min}}
            d_{min} \leqslant  d_{i(t) j(t)}, \forall  i,j\in\mathcal{U}, t\in T.
        \end{equation}
        Here, $d_{min}$ represents the safe distance to avoid collisions among UAVs.}
        
        \textcolor{black}{ 
        However, the fact that UAV $j$ is within the communication range of UAV $i$ 
        does not necessarily represent that $e(i,j)=1$. 
        The reasons are that UAV $j$ may be malicious, and the limited resources of UAV $i$ 
        can only support the communication with a certain number of UAVs. 
        Specially, $\mathfrak{q}$  UAVs are selected to establish effective links, 
        combining the trust value of UAVs and the distance from UAV $i$. 
        The set of the $\mathfrak{q}$ UAVs is represented as $\Gamma_{i(t)}$,
        which is dynamically adjusted according to the network topology.}

\subsection{Attack Model}
Due to the unsecure environment, UAVs may be vulnerable to attacks and 
become malicious nodes.
\textcolor{black}{Hence, it is necessary to model the attack and then 
analyze the impact of attacks on the network. 
Generally, most attacks destroy the network via specific standards, 
such as prioritizing attacks on important nodes.  
Therefore, we design a deliberate attack model based on the node ranking, 
in which UAVs with higher importance are attacked in priority {\cite{attacks, He_Routing}}. 
In this way, the proposed approach can be more effectively applied to defend against 
such attacks and reduce the total delay caused by the malicious activities.}

\textcolor{black}{
Considering the node degree $\zeta_{i(t)}$ of UAV $i$ and the 
weight $I_{i(t)j(t)}$ of link $e_{i,j}$, the importance $\varLambda_{i(t)}$ 
of UAV $i$ at time slot $t$ is calculated as 
\begin{equation}
    \begin{aligned}
    \varLambda _{i(t)} &=  \zeta_{i(t)}  +  \\
    &\sum_{{j} \in \Gamma_{i(t)}}I_{i(t)j(t)}\cdot(1\!-\!\frac{\zeta_{j(t)} \!-\! 1}{\zeta_{i(t)}\! + \!\zeta_{j(t)} \!- \!2}), 
    t \in T,
    \end{aligned}
\end{equation}
in which
\begin{equation}
    \zeta_{i(t)}=\sum_{j \in \mathcal{U}} e(i, j), t \in T,
\end{equation}
and
\begin{equation}
    I_{i(t)j(t)}={Z}\cdot\frac{2}{m+2}.
\end{equation}
Wherein, $Z=(\zeta_{i(t)}\! -\! m \!- \!1)\cdot(\zeta_{j(t)}\! - \!m \!- \!1)$ indicates the 
connectivity ability of link $e_{i,j}$, and $m$ represents the number of 
triangles containing link $e_{i,j}$ in the UAV network topology.}

\textcolor{black}{Besides, we utilize $\mathtt{f}$ and $\mathcal{U}_\mathtt{f}$ to
denote the number and set of malicious UAVs, respectively.}
When malicious UAVs are recognized, they are isolated from the network, 
leading to the unavailability of original routing paths.
Therefore, a new transmission path should be found for recovering 
the routing, as shown in Fig. {\ref{fig:Network_of_UAV}}. 

\vspace{-0.2cm}
\subsection{Communication  Model}
\textcolor{black}{
The path loss $L_{i(t) j(t)}$ of the communication channel between UAVs $i$ and $j$
is approximated by a free space path loss model {\cite{channel_model_PL}}, 
which remains constant within time slot $t$, i.e.,
\begin{equation}{\label{PLU}}
    L_{i(t) j(t)}=10^{\vartheta \lg d_{i(t) j(t)}+ \lg (\frac{4\pi f}{c})^{2}},
\end{equation} 
in which $\vartheta$ indicates the path loss exponent, 
$f$ represents the carrier frequency, and $c$ is the speed of light.}

\textcolor{black}{
Meanwhile, the signal-to-noise ratio $\Psi _{i(t) j(t)}$ in the channel of UAVs $i$ and $j$ 
at time slot $t$ is 
\begin{equation}
    \Psi _{i(t) j(t)}=\frac{\mathcal{P}^{tr}_{i(t) j(t)}}{\sigma^{2}_{i(t) j(t)}{L_{i(t) j(t)}}},
\end{equation}
where $\mathcal{P}^{tr}_{i(t) j(t)}$ and $\sigma^{2}_{i(t) j(t)}$ represent the transmission power 
and noise power, respectively. 
Accordingly, based on Shannon theory {\cite{Shannon_theory}}, 
the transmission rate $G_{i(t) j(t)}$ from UAV $i$ to UAV $j$ at time slot $t$ is 
\begin{equation}{\label{Channel-Rate}}
    G_{i(t) j(t)}=B_{i(t) j(t)}\log_{2}(1+{\Psi _{i(t) j(t)}}),
\end{equation}
where $B_{i(t) j(t)}$ indicates the channel bandwidth between UAVs $i$ and $j$ 
at time slot $t$. Besides, due to limited UAV-to-UAV communication resources, 
the total amount of demands passing $e(i, j)\in \mathcal{E}$ at time slot $t$ should satisfy
\begin{equation}\label{R_L}
   \sum_{r\in R} \eta_{i(t) j(t)}^{r} \mathcal{L}^{r} \leqslant \tau G_{i(t) j(t)} ,\forall  i, j \in \mathcal{U},t \in T, 
\end{equation}
in which $\eta_{i(t) j(t)}^{r}$ is a binary variable denoting whether demand $r$ 
passes the link $e(i, j)$ at time slot $t$,
\begin{equation}
    \eta_{i(t)j(t)}^{r}\!=\!
    \left\{\!
        \begin{aligned}
            &1, \text{ if }  \!r\!  \text { is transmitted via } \!e(i,\! j)\! \text { in time slot }\! t,\\
            &0, \text{ otherwise}.
        \end{aligned}
    \right.
\end{equation}
}
\vspace{-0.5cm}

\subsection{Energy Consumption}

The energy consumption of UAVs is composed of two parts: the communication energy 
consumption of receivers and transmitters, 
and the energy required for UAV movements {\cite{commi_energy,Energy-2}}.
\subsubsection{Communication Consumption}
For UAV $i$, $E^{re}_{i(t)}$ denotes the total energy consumption 
in time slot $t$ for receiving the demand $r\in R$ 
from another UAV $j$, i.e.,  
\begin{equation}
        E^{re}_{i(t)}=\sum_{r\in R}\sum_{j\in \mathcal{U}_\mathtt{s} \cup \mathcal{U}_\mathtt{r}}\mathcal{L}^r \cdot E_{\mathrm{elec}} \cdot \eta^{r}_{j(t) i(t)},
 \end{equation}
and the transmission energy $E^{tr}_{i(t)}$ is
\begin{equation}
E^{tr}_{i(t)}\!=\!\sum_{r\in R}\sum_{j\in \mathcal{U}_\mathtt{r} \cup \mathcal{U}_\mathtt{d}} \mathcal{L}^r  \cdot
(E_{\mathrm{elec}}\!+\!\xi _{fs}  \cdot d_{i(t)j(t)}^2)\cdot \eta^{r}_{i(t) j(t)},
\end{equation}
where $\mathcal{L}^{r}$ and ${E_{\mathrm{elec}}}$ indicate the bit length of demand $r$ and 
the energy consumption of radio electronic devices per bit, respectively {\cite{commi_energy}}. 
$\xi_{fs}$ represents the energy consumption factor of the power amplifier, corresponding to the proposed free space model in (\ref{PLU}).
Besides, $d_{i(t)j(t)}$ denotes the distance between UAVs $i$ and $j$ at time slot $t$.
Then, the communication energy consumption of UAV $i$ 
at time slot $t$ is
\begin{equation}
    E^{com}_{i(t)}=E^{re}_{i(t)}+E^{tr}_{i(t)}.
\end{equation}
\subsubsection{Mobility Consumption}
The mobility energy consumption enables UAVs to hover and move in horizontal
and vertical directions, ensuring that UAVs can operate normally.
\textcolor{black}{According to {\cite{P_0_i}}, in time slot $t$, 
the mobility energy consumption model can be represented as
\begin{equation}
    \begin{aligned}
    E^{mov}_{i(t)}&\!=\!\int_{0}^{\tau} \mathcal{P}^{hov}_{i(t)} \,dt  \\
    &\!+\!\frac{M_{i}\left(\bm{\nu}_{i(t+\tau)}^{2}\!-\!\bm{\nu}_{i({t+0})}^{2}\right)}{2}\! +\!g M_{i} \Delta z_{i(t)},
    \end{aligned}
\end{equation}
where $M_{i}$ and $g$ are the mass of UAV $i$ and the acceleration of 
gravity, respectively. 
$\bm{\nu}_{i(t)}$ is the speed of UAV $i$, and $\Delta z_{i(t)}=z_{i({t+\tau})}-z_{i({t+0})}$ 
is the movement distance against gravity during time slot $t$.
$\mathcal{P}^{hov}_{i(t)}$ represents the hover power for UAV $i$ in the instantaneous time $t$, i.e.,
\begin{equation}
    \begin{aligned}
    &\mathcal{P}^{hov}_{i(t)} \!=\!\mathcal{P}^{hov}_{b} \left( \! 1\!+\!\frac{3\bm{\nu}^{2}_{i(t)}}{U_{tip}^{2}} \! \right) \!+\! \\
    &\mathcal{P}^{hov}_{I} \! \left( \!\sqrt{\left(1 \!+\!\frac{\bm{\nu}^{4}_{i(t)}}{4v_0^{4}}\right)}\!-\!\frac{\bm{\nu}^{2}_{i(t)}}{2v_{0}^{4}} \!\right)^{\frac{1}{2}}
    \!+\!{\frac{1}{2}}\partial_{0} \jmath_{0} \imath_{0} A_{0} \bm{\nu}^{3}_{i(t)},
    \end{aligned}
\end{equation}
where $\mathcal{P}^{hov}_{b}$ and $\mathcal{P}^{hov}_{I}$ indicate 
the blade profile power and induced power at hovering states in {\cite{P_0_i}}, respectively.}
$U_{tip}$ and $v_{0}$ denote the tip speed of rotor blades and 
average rotor induction speed hovering, respectively.
$\partial _{0}$,  $\jmath _{0}$, $\imath _{0}$, and $A_{0}$ 
represent the airframe drag ratio, air density, rotor solidity,
and rotor disk area, respectively. 

Consequently, the total energy consumption of UAV $i$ at time slot $t$ is
\begin{equation}
    E^{con}_{i(t)}=E^{com}_{i(t)} + E^{mov}_{i(t)}.
\end{equation}

Besides, the total energy consumption cannot violate the energy capacity 
$E^{max}_{i(t)}$ of UAV $i$ at time slot $t$, i.e., 
\begin{equation}\label{energy}
    E^{con}_{i(t)}  \leq \beta E^{max}_{i(t)}, \forall i \in {\mathcal{U}}, t\in T,
\end{equation}
where $\beta$ is the empirical power efficiency.

It is set within a rationale value, taking into account the recommended discharge of depth 
for the specific UAV battery and user safety requirements  {\cite{DoD,DOD-080}}.

\vspace{-0.2cm}
\subsection{Delay Model}
The end-to-end routing delay mainly consists of the queue delay 
and transmission delay on the multi-hop path from the source UAV to destination UAV.

\subsubsection{Queue Delay ($\mathcal{T}^{que,r}$)}
Since UAV networks are characterized by high mobility and
unstable data traffic fluctuations,
we introduce a queue buffer for each UAV to alleviate the network congestion.
In particular, the amounts of demands received and transmitted by UAV $i$ 
are denoted as $\varepsilon_{i(t)}^{re}$ and $\varepsilon_{i(t)}^{tr}$, respectively, i.e.,
\begin{align}{\label{re_tr}}
    \left\{\begin{aligned}
    &\varepsilon_{i(t)}^{re}=\sum_{r\in R} \sum_{j \in \mathcal{U}_\mathtt{s}\cup \mathcal{U}_\mathtt{r} } \eta^r_{j(t)i(t)},\\
    &\varepsilon_{i(t)}^{tr}=\sum_{r\in R} \sum_{j \in \mathcal{U}_\mathtt{r} \cup \mathcal{U}_\mathtt{d} } \eta^r_{i(t)j(t)},
    \end{aligned}\right.\quad t \in T.
\end{align}
Further, the transformation of queue length $\Delta C_{i(t)}$ 
from time slot $t-1$ to $t$ of UAV $i$ is equal to the difference between
the numbers of received demands $\varepsilon_{i(t-1)}^{re}$ 
and transmitted demands $\varepsilon_{i(t-1)}^{tr}$ at time slot $t-1$, i.e.,
\begin{equation}
    \Delta C_{i(t)}=\varepsilon_{i(t-1)}^{re}-\varepsilon_{i({t-1})}^{tr}.
 \end{equation}
Then, for UAV $i$, $C_{i}^{max}$ and $C_{i(t)}$ 
denote the maximum queue capacity and the total amount of 
queued demands (i.e., the queue length), respectively, which should satisfy
\begin{align}{\label{queue_capacity}}
    \left\{\begin{aligned}
    &0\leq C_{i(t)}\leq C_{i}^{max},\\
    &0\leq \varepsilon_{i(t)}^{re}\leq C_{i}^{max},\\
   & 0\leq \varepsilon_{i(t)}^{tr}\leq C_{i(t)}.
    \end{aligned}\right.\end{align}
Additionally, $C_{i(t)}$ is calculated by adding $\Delta C_{i(t)}$ to 
the queue length at time slot $t-1$, i.e.,
\begin{equation}
    C_{i(t)}=C_{i({t-1})}+\Delta C_{i({t})}.
\end{equation}

\vspace{-0.2cm}
\textcolor{black}{Based on the above analysis, 
when there exists demand $r \in R$ arriving at UAV $i$ with $C_{i({t})}$, 
the queue delay of demand $r$ for transmitting to next-hop UAV $j$ is  
\begin{equation}{\label{queue-delay}}
    \mathcal{T}^{que,r}_{i(t)}=\sum\limits_{r\in \mathcal{C}_{i(t)}}\frac{\eta^{r}_{i(t) k(t)} \mathcal{L}^r}{G_{i(t)k(t)}}, k\in \Gamma_{i(t)}, 
 \end{equation}
where $\mathcal{C}_{i(t)}$ is the queued demand set of UAV $i$ at time slot $t$, 
and $|\mathcal{C}_{i(t)}|=C_{i(t)}$.     
Besides, to maintain  the orderly processing of demands, the queuing model leverages first-in-first-out rules.}

\subsubsection{Transmission Delay ($\mathcal{T}^{tr,r}$)}
\textcolor{black}{
The one-hop transmission delay of demand $r$ 
from UAV $i$ to UAV $j$ at time slot $t$ is 
\begin{equation}{\label{trans-delay}}
    \mathcal{T}^{tr,r}_{i(t) j(t)}= \frac{\mathcal{L}^{r}}{G_{i(t)j(t)}} \eta^{r}_{i(t) j(t)}, r\in R.
\end{equation}}

\textcolor{black}{
Hence, the sum of one-hop delay for transmitting demand ${r}$ 
from UAV $i$ to UAV $j$ at time slot $t$ is 
\begin{equation}{\label{delay}}
    \mathcal{T}^{r}_{i(t) j(t)}= \mathcal{T}^{tr,r}_{i(t) j(t)} +  \mathcal{T}^{que,r}_{i(t) j(t)}, t \in T,
\end{equation}
in which the value of $\mathcal{T}^{r}_{i(t) j(t)}$ cannot 
exceed the maximum tolerated delay $\mathcal{T}_{one}^{max}$ for one-hop, i.e.,
\begin{equation}{\label{delay_constraint}}
    \mathcal{T}^{r}_{i(t) j(t)} \leq \mathcal{T}_{one}^{max}.
\end{equation}
When the routing path $\mathsf{P}^{r}_{\mathtt{sd}}$ for transmitting demand $r$ 
from source UAV $\mathtt{s}$ to destination UAV $\mathtt{d}$ 
is determined, the end-to-end delay $\mathcal{T}^{r}$ can 
be calculated as 
\begin{equation}
    \mathcal{T}^{r}=\underset{t\in T}{\sum} \underset{{e(i,j)\in \mathsf{P}_{\mathtt{sd}}^r}}{\sum}\mathcal{T}^{r}_{i(t)}.
\end{equation}
}
 
\vspace{-1cm}
\subsection{Problem Formulation}
\subsubsection{Flow Constraints} 
\textcolor{black}{
The demand $r$ from source UAV $ \mathtt{s} \in \mathcal{U}_\mathtt{s}$ can 
only be received by one another UAV 
$j$, i.e.,
\vspace{0.1cm}
\begin{equation}\label{F1}
    \sum_{j \in \mathcal{U}_\mathtt{r} \cup \mathcal{U}_\mathtt{d}}\eta^{r}_{\mathtt{s}(t) j(t)}=1, 
    \forall r \in R, t \in T.
\end{equation} 
Recall that $\eta^{r}_{\mathtt{s}(t) j(t)} \in \{0,1\}$ denotes 
whether demand $r$ via link $e({\mathtt{s}, j})\in \mathcal{E}$, 
1 if passing and 0 otherwise.
With respect to a middle UAV $j \in \mathcal{U}_\mathtt{r}$, the flow conservation 
should be satisfied:
\begin{align}{\label{F2}}
    \!\left\{\!\begin{aligned}
            &\sum_{i  \in \mathcal{U}_{\mathtt{s}}\cup \mathcal{U}_\mathtt{r}} 
            \eta^{r}_{{i(t)} {j(t)}}\!= \!
            \sum_{i  \in \mathcal{U}_\mathtt{r} \cup \mathcal{U}_{\mathtt{d}}}
            \eta^{r}_{{j(t)} {i(t)}}\!+\!
            \eta^{r}_{{j(t)} {j(t+1)}},\\
            &\qquad\qquad\qquad\qquad\quad\forall r \in R,  j\in \mathcal{U}_\mathtt{r}, t=1,\\
            &\sum_{i  \in \mathcal{U}_{\mathtt{s}}\cup \mathcal{U}_\mathtt{r}} 
            \eta^{r}_{{i(t)} {j(t)}} + \eta^{r}_{{j(t-1)} {j(t)}}  \!=  \!
            \sum_{i  \in \mathcal{U}_\mathtt{r} \cup \mathcal{U}_{\mathtt{d}}}
            \eta^{r}_{{j (t)} {i(t)}} \!+\! \\
            &\qquad\eta^{r}_{{j(t)} {j(t+1)}},
            \forall r \in R,  j\in \mathcal{U}_\mathtt{r}, t\in \{2,\cdots,T-1\},\\
            &\sum_{i  \in \mathcal{U}_{\mathtt{s}}\cup \mathcal{U}_\mathtt{r}} 
            \eta^{r}_{{i(t)} {j(t)}}+\eta^{r}_{{j(t-1)} {j(t)}}\!= \!
            \sum_{i  \in \mathcal{U}_\mathtt{r} \cup \mathcal{U}_{\mathtt{d}}}
            \eta^{r}_{{j(t)} {i(t)}},\\
            &\qquad\qquad\qquad\qquad\quad\forall r \in R,  j\in \mathcal{U}_\mathtt{r}, t=T.
        \end{aligned}
            \right.
\end{align}
Besides, a demand can select only one routing path, i.e.,
\begin{align}{\label{F3}}
    \left\{\begin{aligned}
        &\sum_{j  \in \mathcal{U}_\mathtt{r} \cup \mathcal{U}_{\mathtt{d}}}
        \eta^{r}_{{i}(t) {j}(t)}+
        \eta^{r}_{{i(t)} {i(t+1)}}= 1,\\
        &\qquad\qquad \forall r \in R,  i\in \mathcal{U}_\mathtt{r}, t\in \{1,\cdots,T-1\},\\
        &\sum_{j  \in \mathcal{U}_\mathtt{r} \cup \mathcal{U}_{\mathtt{d}}}
        \eta^{r}_{{i(t)} {j(t)}}= 1, \forall  r \in R,  i\in \mathcal{U}_\mathtt{r}, t=T.
    \end{aligned}
    \right.
 \end{align}
Similar to the source UAVs, 
the destination UAV $\mathtt{d} \in \mathcal{U}_\mathtt{d}$ 
can only receive demand $r$ from one another UAV $i$, i.e.,
\begin{equation}\label{F4}
    \sum_{i \in \mathcal{U}_\mathtt{s} \cup \mathcal{U}_\mathtt{r}}\eta^{r}_{i(t) \mathtt{d}(t)}=1, 
    \forall r \in R, t \in T.
\end{equation} }
\subsubsection{Problem Formulation}
The objective is to minimize the total end-to-end delay of networks 
with attacked UAVs that may become malicious nodes, 
and the corresponding optimization problem is formulated as
\begin{equation}{\label{optimal}}
    \begin{aligned}
    \mathscr{P}0:\;\underset{{\boldsymbol{\eta}}}{\textrm{min}}\;
    \! &\underset{r \in R} {\sum}
    \mathcal{T}^{r}\\
       \textrm{s.t.}\;
       &(\ref{distance_min}), (\ref{distance_max}), (\ref{R_L}),(\ref{energy}), 
       (\ref{queue_capacity}), (\ref{delay_constraint}), (\ref{F1})\!-\!(\ref{F4}),\\
       &\eta_{i(t)j(t)}^{r} \in \{0,1\}, \\  
       &\forall t\in T, e(i,j)\! \in \mathcal{E},i,j\in \mathcal{U}, i,j \notin \mathcal{U}_\mathtt{f}, r\in R,
    \end{aligned}
\end{equation}
where $\boldsymbol{\eta}=\{\eta_{i(t)j(t)}^{r},\forall t\in T, e(i,j)\in \mathcal{E}, 
i,j \in \mathcal{U}, i,j \notin \mathcal{U}_\mathtt{f}, r \in R\}$, in which 
$i,j \notin \mathcal{U}_\mathtt{f}$ represents that demand $r\in R$ 
avoids being transmitted by malicious UAVs in the network.
It is observed that $\mathscr{P}0$ is in the form of ILP and 
NP-hard to deal with {\cite{Lancia2018}}. 
Therefore, in the next section we propose the feasible solutions. 
\section{Algorithm Design \label{sec:Reinforcement-learning}}
In this section, to realize the security in the UAV network 
with malicious nodes, the BTMM is proposed. Further, for 
effectively solving $\mathscr{P}0$, we reformulate it into 
a Dec-POMDP and propose an MADDQN-based algorithm 
to make dynamic routing decisions.
\subsection{Blockchain-based Trust Management Mechanism}
    Distributed trust mechanisms enable UAVs to autonomously evaluate the reliability of neighbor UAVs 
    without relying on a central authority. 
    This approach enhances the scalability and fault tolerance, especially for dynamic UAV networks {\cite{DTM}}.
Since the consortium blockchain is decentralized and jointly maintained 
by multiple pre-selected nodes, it can be applied for UAVs to create 
the secure and reliable communication network.
In detail, the malicious behaviors of UAVs are analyzed, 
and we design a mechanism to evaluate the trust value of UAV nodes.
Besides, a consensus UAV update mechanism is proposed to improve 
the security of the PBFT consensus algorithm, 
which isolates UAVs with trust values below a safety threshold. 
\subsubsection{Node Trust Evaluation Mechanism}
The malicious behaviors of UAVs are analyzed in two aspects. In particular, 
when the malicious UAVs receive the demand, they may drop it or do not 
transmit it following the specific routing path. 
Hence, considering the delivery rate and the correctness of transmission paths, 
the trust evaluation mechanism for each UAV is designed as follows.
\begin{itemize}
    \item {\em Delivery rate evaluation}: The value of the delivery rate is applied 
    to evaluate the malicious UAVs with behaviors rejecting to transmit 
    received demands, which is represented as the ratio of 
    total transmitted demands $\sum_{k=1}^{t}\varepsilon^{tr}_{i(k)}$ to total received
    demands $\sum_{k=1}^{t}\varepsilon^{re}_{i(k)}$ by UAV $i$ at time slot $t$, i.e.,
    \begin{equation}{\label{equ:dr}}
        \mathbb{T}_{i(t) }^{dr}\!=\!\frac{\sum\limits_{k=1}^{t}\! \varepsilon^{tr}_{i(k)}}{\sum\limits_{k=1}^{t}\! \varepsilon^{re}_{i(k)}}, \forall i \in \mathcal{U},\sum_{k=1}^{t} \!\varepsilon^{re}_{i(k)} \!> \!0, t \in T.
    \end{equation} 
    Besides, $\sum_{k=1}^{t} \!\varepsilon^{re}_{i(k)}\!=\!0$ denotes UAV $i$ does not receive demands, 
    and $\mathbb{T}_{i(t) }^{dr}$ remains the initial value.

    \item {\em Transmission path evaluation}: The malicious UAVs typically 
    do not follow the specified path for transmitting demands
    to the next-hop UAV, which results in increasing 
    transmission delay {\cite{delay_9312485}}. Therefore, we evaluate 
    the correctness of transmission paths to partly assess whether 
    the UAV is malicious. When UAV $j$ receives a demand $r$, it checks 
    whether the data is transmitted by UAV $i$ following 
    specified paths. If not, the normal UAV $j$ records 
    the relevant abnormal information of UAV $i$ and then uploads it to the consensus UAVs 
    as a basis for calculating node trust values. 
    When the UAV is not recorded, it is defaulted as a safe node. 
    In particular, we can calculate the evaluation value of UAV $i$ in terms of the 
    transmission path at time slot $t$ by 
    \begin{equation}{\label{equ:tp}}
        \mathbb{T}_{i(t)}^{tp}=1-\frac{{\sum\limits_{k=1}^{t}}\mathfrak{B}_{i(k)}}{{\sum\limits_{k=1}^{t}}\mathbb{B}_{i(k)}}, \forall i \in \mathcal{U},{\sum\limits_{k=1}^{t}}\mathbb{B}_{i(k)}\!>\!0, t \in T,
    \end{equation}
    where $\sum_{k=1}^{t}\mathfrak{B}_{i(t)}$ and $\sum_{k=1}^{t}\mathbb{B}_{i(t)}$ indicate the total numbers of non-specified and 
    total transmission paths of UAV $i$ at time slot $t$, respectively. In addition, when $\sum_{k=1}^{t}\mathbb{B}_{i(t)}=0$, 
    $\mathbb{T}_{i(t)}^{tp}$ remains unchanged at the initial value.
\end{itemize}
    
\begin{algorithm}[t]
    \caption{\label{Alg:trust-weight}{Adaptive Weights based Trust Evaluation Process}}
    \begin{algorithmic}[1]
        \STATE {\textbf{Initialization:}  $\mathbb{T}_{i(1)}\! = \!1, \mathbb{T}^{dr}_{i(1)}\!=\!1,  \text{and } \mathbb{T}^{tp}_{i(1)}\!=\!1, \forall i\! \in \!\{\!1,...,N\!\}$}.
        Initialize the values of the demand delivery probability $p_1$ 
        and the correct path selection probability $p_2$, respectively.
        \FOR {each time step $t$}
            \FOR{{each UAV $i$}}
                \STATE {Update $\mathbb{T}_{i(t)}^{dr}$ via ({\ref{equ:dr}}).}
                \STATE {Update $\mathbb{T}_{i(t)}^{tp}$ via ({\ref{equ:tp}}).}
                \STATE { Calculate $\psi _{i(t)}^0, \psi _{i(t)}^1$, and $\psi _{i(t)}^2$ by ({\ref{Adaptive_weights}}).}
                \STATE { Compute $\mathbb{T}_{i(t)}$ according to ({\ref{equ:trust-total}}).}
                \IF {$\mathbb{T}_{i(t)} < \mathbb{T}_{thr}$}
                    \STATE { Flag UAV $i$ as  malicious.}
                 \ENDIF 
            \ENDFOR 
         \ENDFOR 
\end{algorithmic}
\end{algorithm}

Considering the above two factors, the comprehensive trust value of UAV 
$i$ at time slot $t+1$ is calculated as 
\begin{equation}
    {\label{equ:trust-total}}
    \begin{aligned}
    \mathbb{T}_{i(t+1)} =\psi _{i(t)}^0\mathbb{T}_{i(t)}+ \psi_{i(t)}^1 \mathbb{T}^{dr}_{i(t)}& +\psi_{i(t)}^2 \mathbb{T}^{tp}_{i(t)}, \\
    & \forall  i \in \mathcal{U}, t\in T.      
    \end{aligned}
\end{equation}
Wherein, $\mathbb{T}_{i(1)}\! = \!1, \mathbb{T}_{i(1)}^{dr}\!=\!1,  \text{and } \mathbb{T}^{tp}_{i(1)}\!=\!1$ 
indicate the initial trust values for UAV $i$. 
$\psi _{i(t)}^0$, $\psi _{i(t)}^1$, and $\psi _{i(t)}^2$ are hyperparameters, 
denoting the weights of $\mathbb{T}_{i(t)}$,
 $\mathbb{T}^{dr}_{i(t)}$, and $\mathbb{T}^{tp}_{i(t)}$, respectively, and $\psi_{i(t)}^0\!+\!\psi _{i(t)}^2\!+\!\psi_{i(t)}^2\!=\!1$.

    Moreover, considering that $\mathbb{T}^{dr}_{i(t)}$ and  $\mathbb{T}^{tp}_{i(t)}$ are directly related 
    to the trust value evaluation, we propose the adaptive method to determine the weights of trust values,  detailed as 
\begin{align}
    \left\{\begin{aligned}{\label{Adaptive_weights}}
        \psi _{i(t)}^0&\!=\! 0.5 \times \frac{\mathbb{T}_{thr}}{\mathbb{T}_{i(t)}}, \\
        \psi _{i(t)}^1 &\!=\! \frac{(1\!-\!\psi _{i(t)}^0)\cdot(1\!-\!\mathbb{T}^{dr}_{i(t)})}{2\! -\! \mathbb{T}^{dr}_{i(t)}\! +\! \mathbb{T}^{tp}_{i(t)}}, 0\leq\! \mathbb{T}^{dr}_{i(t)} \!+ \!\mathbb{T}^{tp}_{i(t)}\!\leq 2,\\
        \psi _{i(t)}^2 &\!=\! \frac{(1\!-\!\psi _{i(t)}^0)\cdot(1\!-\!\mathbb{T}^{tp}_{i(t)})}{2 \!-\! \mathbb{T}^{dr}_{i(t)}\! + \!\mathbb{T}^{tp}_{i(t)}}.
    \end{aligned}\right.
\end{align}
When $\mathbb{T}^{dr}_{i(t)}\! +\! \mathbb{T}^{tp}_{i(t)}\!=\!2$, $\psi _{i(t)}^1 \!=\!\psi _{i(t)}^2\!=\!\frac{(1\!-\!\psi _{i(t)}^0)}{2}$.
$\mathbb{T}_{thr}$ can be determined by the security requirements {\cite{trust_parameters}}.

Algorithm {\ref{Alg:trust-weight}} indicates the trust evaluation 
process of the adaptive weight method. 
For each UAV $i$, initialize $\mathbb{T}_{i(1)}\! = \!1, \mathbb{T}^{dr}_{i(1)}\!=\!1,  \text{and } \mathbb{T}^{tp}_{i(1)}\!=\!1$ as 1. 
Additionally, initialize the probabilities associated with the behaviors of malicious UAVs. 
Specifically, the demand delivery probability and the correct path selection probability are denoted as $p_1$ and $p_2$, respectively.
At each time step $t$, update $\mathbb{T}^{dr}_{i(t)}$ via (33) 
and $\mathbb{T}^{tp}_{i(t)}$ via (34).
Calculate $\psi _{i(t)}^0, \psi _{i(t)}^1$, and $\psi _{i(t)}^2$ according to (36).
Then, compute $\mathbb{T}_{i(t)}$ based on (35).
If $\mathbb{T}_{i(t)}$ is less than $\mathbb{T}_{thr}$, flag UAV $i$ as malicious.

\subsubsection{Trust-based PBFT Algorithm}
\begin{figure*}[t]
    \centering
    \includegraphics[width=0.86\linewidth]{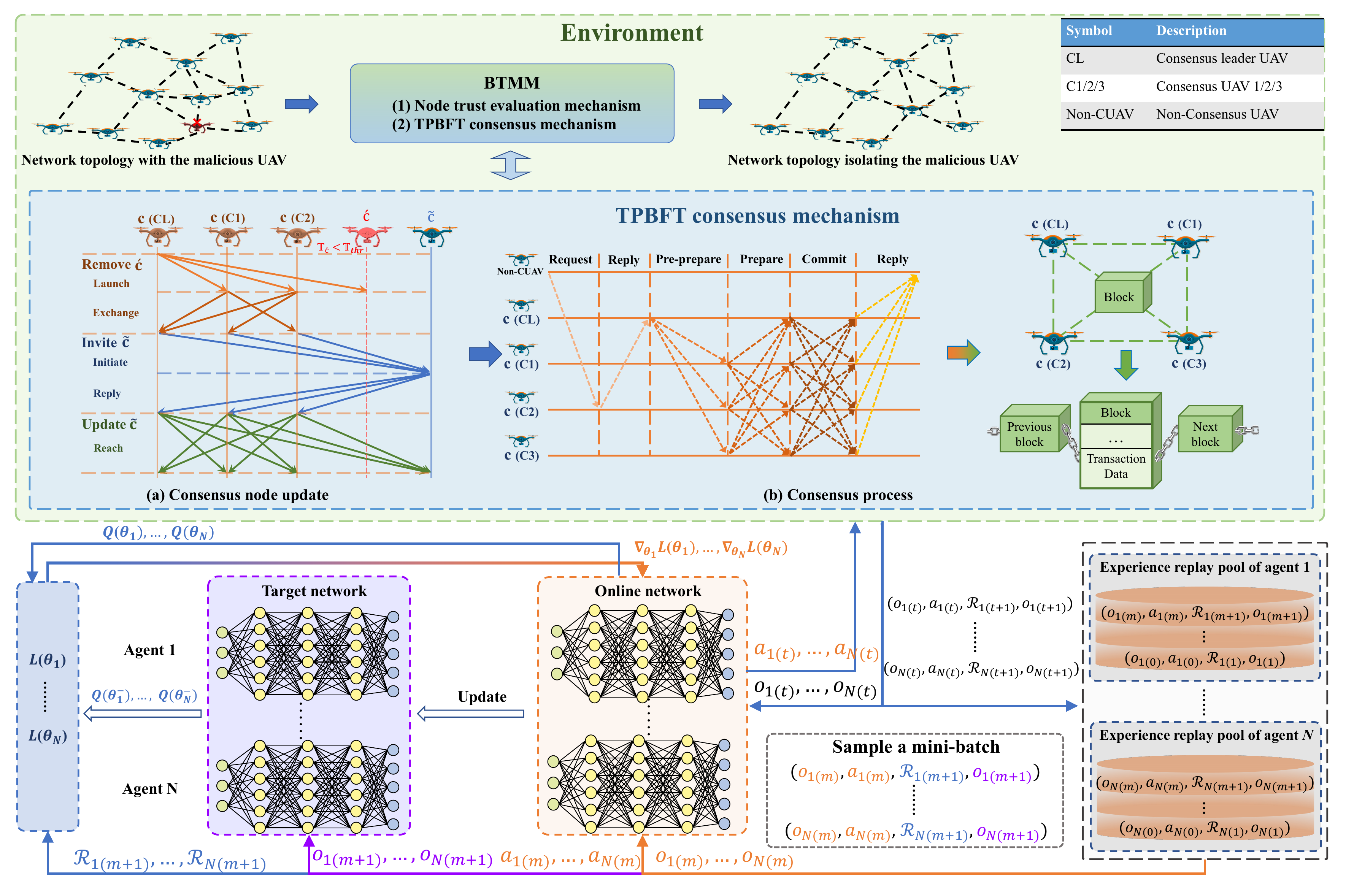}
    \textcolor{black}{\caption{\label{fig:consensus_update} BTMM-MADDQN algorithm 
    framework for trusted routing in the UAV network.}}
    \vspace{-0.2cm}
\end{figure*}

{\label{Consensus_Mechanism}}
To reach the consensus in distributed UAV networks with malicious nodes, 
we introduce the PBFT algorithm, which can tolerate less than 1/3 
fault nodes \cite{PBFT_repu}. 
In this work, to further ensure the security of PBFT processes,
we propose the trust-based PBFT (TPBFT) algorithm, 
which updates the consensus UAV set $\mathcal{U}_\mathtt{c}$, 
as illustrated in Algorithm {\ref{Alg:consensus}}. 
In detail, to initialize the network, we select $3n+1$ UAVs with the largest trust values 
as consensus node $\mathtt{c} \in \mathcal{U}_\mathtt{c}$, 
in which the UAV with highest trust value is selected as the leader, and the candidate 
consensus UAV set is denoted as $\mathcal{U}_{\tilde{{\mathtt{c}}}}$ (line {\ref{Alg1_1}}).
With $\mathcal{U}_\mathtt{c}$ identified, the consensus process for transactions 
in Fig. {\ref{fig:consensus_update}} is performed and primarily includes three stages: \textit{pre-prepare}, 
\textit{prepare}, and \textit{commit} (line {\ref{Alg1_3}}).
In particular, the non-consensus UAVs upload transactions with recorded  
information to the nearest consensus UAV, which relays it to the leader consensus UAV
for verification, and then the leader starts the three-stage processes
among all consensus UAVs as follows.

    \begin{itemize}
    
    \item \textcolor{black}{Firstly, the leader packs the validated transaction into a block and then broadcasts the \textit{pre-prepare} message 
    $\langle \langle \textit{pre-prepare}, \mathfrak{k}, \mathfrak{n}, \mathfrak{d} \rangle_{\mathfrak{a } _\mathfrak{p} }, \mathfrak{m} \rangle$ 
    to non-leader consensus UAVs. Wherein, \(\mathfrak{k}\) is the current view,  \(\mathfrak{n}\) is the sequence number, 
    \(\mathfrak{m}\) is the original request,  and the digest of \(\mathfrak{m}\) is \(\mathfrak{d}\).
    The leader attaches its digital signature \({\mathfrak{a }_\mathfrak{p} }\) and message authentication codes (MACs) to ensure the authenticity.}
    \item \textcolor{black}{In the \textit{prepare} stages, each non-leader consensus UAV verifies the received \textit{pre-prepare} message 
    by checking the signature and MACs of the leader and the consistency of \(\mathfrak{d}\) with \(\mathfrak{m}\). 
    After successful verification, consensus UAV \(\mathtt{c}\) broadcasts a \textit{prepare}  message 
    \(\langle \textit{prepare}, \mathfrak{k}, \mathfrak{n}, \mathfrak{d}, \mathtt{c} \rangle_{{\mathfrak{a } _\mathtt{c} }}\)  
    to other consensus UAVs. Furthermore, each consensus UAV compares the received \textit{prepare} messages with 
    the original \textit{pre-prepare} message to ensure the consistency of (\(\mathfrak{k}, \mathfrak{n}, \mathfrak{d}\)). 
    The consensus requires more than \( 2n\) valid \textit{prepare} signatures,
    where \(n\) is the maximum number of faulty consensus UAVs.}
    \item \textcolor{black}{In the \textit{commit} phase, if the consensus UAV receives more 
    than or equal to $2n$ valid \textit{prepare} messages, they broadcast \textit{commit} messages 
    to the other consensus UAVs for verification. If each consensus UAV receives $2n+1$ \textit{commit} 
    messages, the consensus is reached.}
\end{itemize}

\color{black}
\begin{algorithm}[t]
    
    \caption{TPBFT Algorithm \label{Alg:consensus}}
\begin{algorithmic}[1]
\REQUIRE {UAV set $\mathcal{U}$, security threshold $\mathbb{T}_{thr}$,
 and initial trust value $\mathbb{T}_{i(1)}$ for all $i \in \mathcal{U}$}.
\ENSURE $\mathcal{U}$ and consensus UAV set $\mathcal{U}_\mathtt{c}$.
\STATE \label{Alg1_1} \textbf{Initialization:} $\mathcal{U}$, $\mathbb{T}_{i(1)}$, $\mathbb{T}_{thr}$, $\mathcal{U}_\mathtt{c}$,  
and candidate consensus UAV set $\mathcal{U}_{\tilde{{\mathtt{c}}}}$.
\FOR{$t \in T$}\label{Alg1_2} 
\STATE \label{Alg1_3}The non-consensus UAVs upload transactions for verification, and  
the consensus is reached among all $\mathtt{c} \in \mathcal{U}_\mathtt{c}$ 
via the three-stage processes illustrated in Fig. {\ref{fig:consensus_update}}.
 \IF {$\mathcal{M}$ rounds of consensus processes are finished} \label{Alg1_4} 
            \IF {$\exists \mathtt{c} \in \mathcal{U}_\mathtt{c}$, $\mathbb{T}_{\mathtt{c}(t)}<\mathbb{T}_{thr}$} \label{Alg1_5}
                \STATE Delete UAV $\mathtt{c}$ from  $\mathcal{U}_\mathtt{c}$. 
            \ENDIF\label{Alg1_6}
            \IF {$\forall \mathtt{c} \in \mathcal{U}_\mathtt{c}$, $\mathbb{T}_{\mathtt{c}(t)}>\mathbb{T}_{thr}$} \label{Alg1_8}
                \STATE Delete consensus UAV $\mathtt{c}$ with the minimum trust value from $\mathcal{U}_\mathtt{c}$. 
            \ENDIF\label{Alg1_9} 
            \STATE \label{Alg1_11}  Invite candidate UAV ${\tilde{{\mathtt{c}}}} \in \mathcal{U}_{\tilde{{\mathtt{c}}}}$ 
            with the maximum trust value $\mathbb{T}_{\tilde{{\mathtt{c}}}(t)}$ to 
            $\mathcal{U}_\mathtt{c}$.
    \ENDIF  \label{Alg1_12}
\ENDFOR\label{Alg1_13}
\end{algorithmic}
\end{algorithm}

\color{black}

Further, in Algorithm {\ref{Alg:consensus}}, if $\mathcal{M}$ rounds of consensus processes are finished, 
the process for updating $\mathcal{U}_\mathtt{c}$ is performed (line \ref{Alg1_4}). 
If there exists UAV $\mathtt{c} \in \mathcal{U}_\mathtt{c}$ 
with a lower trust value than the safety threshold $\mathbb{T}_{thr}$, 
it is removed from the $\mathcal{U}_\mathtt{c}$ (lines {\ref{Alg1_5}}-{\ref{Alg1_6}}). 
Otherwise, the consensus UAV $\mathtt{c}$ with the minimum trust value 
is removed from $\mathcal{U}_\mathtt{c}$ (lines {\ref{Alg1_8}-{\ref{Alg1_9}}}).
Then, a new consensus UAV is invited from the candidate set $\mathcal{U}_{\tilde{{\mathtt{c}}}}$ (line {\ref{Alg1_11}}).
\textcolor{black}{
Fig. {\ref{fig:consensus_update}} illustrates the node update process 
when there exists one ($\mathtt{f}=1$) malicious consensus UAV, detailed as follows.}
\begin{itemize}
    \item {\em Remove malicious consensus UAV $\acute{\mathtt{c}}$}: 
    The consensus leader UAV launches the 
    request for updating the set $\mathcal{U}_\mathtt{c}$ and sends the information of the consensus UAV 
    with lower trust values (i.e., UAV $\acute{\mathtt{c}}$) than $\mathbb{T}_{thr}$ 
    to the other UAVs in $\mathcal{U}_\mathtt{c}$. 
    In both the launching and exchanging phases, when consensus UAV 
    $\mathtt{c}\in \mathcal{U}_{\mathtt{c}\backslash \acute{\mathtt{c}}}$ receives $2\mathtt{f}$ 
    messages that UAV $\acute{\mathtt{c}}$ is untrusted, it is deleted from consensus
    set $\mathcal{U}_{\mathtt{c}}$.
    \item {\em Invite candidate consensus UAV $\tilde{{\mathtt{c}}}$}: After removing UAV $\acute{\mathtt{c}}$, 
    the other consensus UAVs send an invitation message to a non-consensus 
    UAV $\tilde{{\mathtt{c}}}$ with the highest trust value as the candidate. 
    When UAV $\tilde{{\mathtt{c}}}$ receives $2\mathtt{f}+1$ invitation messages, it sends
    a reply message to UAV $\mathtt{c}\in \mathcal{U}_{\mathtt{c}}$ 
    for the acceptance of invitation.
    \item {\em Update consensus UAV $\tilde{{\mathtt{c}}}$}:
    When consensus UAVs in $\mathcal{U}_{\mathtt{c}}$ receive $2\mathtt{f}+1$ 
    reply messages, they send the update of UAV $\tilde{{\mathtt{c}}}$ 
    to the other consensus UAVs and inform $\tilde{{\mathtt{c}}}$ 
    that it is updated as a formal consensus UAV of $\mathcal{U}_{\mathtt{c}}$. 
    Therefore, the consensus node update is finished.
\end{itemize}

According to the verified transaction, consensus UAVs 
recalculate the trust value based on the designed 
node trust evaluation mechanism, and isolate the UAVs with values below 
the safety threshold, to ensure the network security for transmitting demands. 
It is noted that trust values are managed by the selected consensus UAVs, 
which can avoid the single-node fault and tolerate 1/3 
malicious routing nodes to construct trusted networks. 
Besides, when the malicious UAV is removed from the network, 
the topology variation puts forward the requirement for
dynamic routing solutions. Accordingly, we further propose the 
DRL-based method to deal with $\mathscr{P}0$, which is reformulated
based on the Dec-POMDP.

\subsection{Dec-POMDP Reformulation for the Trusted Routing Problem}{\label{Dec-POMDP}}
Based on the trusted environment, since each UAV can only 
obtain local information from its neighbor UAVs for the distributed network, 
we reformulate $\mathscr{P}0$ into a Dec-POMDP,
which mainly consists of five parts: agent set $\mathcal{N}$, 
state space $\mathcal{S}$, action space $\mathcal{A}$, reward function $\mathcal{R}$, 
and discount factor $\gamma$. 
\begin{itemize}
    \item {\em Agent set}: $\mathcal{N}=\{1,\cdots, N\}$ represents the 
    set of agents (all UAVs). 
    \item {\em State space}: At the start of each time slot $t$, 
    each agent $i \in \mathcal{U}$ gathers and obtains its own observation $o_{i(t)}$
    from the environment. \textcolor{black}{It is noted that each UAV can only observe the information  
    of neighbor node $k\in \Gamma_{i(t)}$.
    Hence, the observation of agent $i$ is 
    \begin{equation}
        \begin{aligned}
        o_{i(t)}\!=\!\{{\varTheta}_{i(t)}, \mathcal{C}_{i(t)},& {\varTheta}_{k(t)}, \mathcal{C}_{k(t)}, E^{con}_{k(t)}, 
        \mathbb{T}_{k(t)}\}, \\
        &\forall i\in \mathcal{U},  k\in \Gamma_{i(t)}, t\in T,
        \end{aligned}
    \end{equation}
    in which  ${\varTheta}_{i(t)}$ and $\mathcal{C}_{i(t)}$ indicate the location 
    and the queued demand set of UAV $i$, respectively.}
    ${\varTheta}_{k(t)}$, $\mathcal{C}_{k(t)}$, $E^{con}_{k(t)}$ and $\mathbb{T}_{k(t)}$ represent 
    the location, queued demand set,  energy consumption, and trust value of agent $k$ at time slot $t$, respectively.
    The observations of all agents are obtained to aggregate the joint state $\bm s_{t}$
    in time slot $t$, denoted as 
    \begin{equation}{\label{state-all}}
        \bm s_{t}=\{o_{1(t)},o_{2(t)}, \cdots,o_{N(t)}\}, t\in T,
    \end{equation}
    and the state space is indicated as $\mathcal{S}=\{\bm s_{t}, t\in T\}$. 

    \textcolor{black}{
    As introduced in Section III-A, the number of UAVs in $k\in \Gamma_{i(t)}$ does not exceed $\mathfrak{q}$, 
    and these UAVs are dynamically adjusted with the network variation. 
    Additionally, the observation dimension of one neighbor UAV is fixed  according to (37).
    Hence, when there exist UAVs joining or leaving, the dimension of observation $o_{i(t)}$ 
    is stabilized via reselecting UAVs or performing zero-padding operations on the remaining dimensions.}
    \item {\em Action space}: 
    \textcolor{black}{
    According to observation $o_{i(t)}$ at time slot $t$, 
    the action set of agent $i$ is denoted as 
    \begin{equation}
        a_{i(t)}=\{a_{i(t)}^k\}, k\in \Gamma_{i(t)}, t\in T,
    \end{equation}
    where $a_{i(t)}^k$ represents that the next step of UAV $i$ is UAV $k$ 
    at time slot $t$. Besides, the joint action of $N$ agents at time slot $t$ is 
    \begin{equation}{\label{action-all}}
        {\bm a_{t}}=\{a_{1(t)},a_{2(t)},\cdots, a_{N(t)}\},t\in T.
    \end{equation}
    Correspondingly, the joint action space is indicated 
    as $\mathcal{A}=\{\bm a_{t},t\in T\}$.}
    \item {\em Reward}:     \textcolor{black}{
    $\mathcal{R}_{i({t+1})}$ indicates the instantaneous reward of 
    agent $i$ performing action $a_{i(t)}$ at time slot $t$. }
    The objective in $\mathscr{P}0$ is to minimize the 
    total end-to-end delay cost of all demands, for which all agents cooperate 
    to maximize their contributions, and then each agent obtains the same reward $\mathcal{R}_{t+1}$. 
    \textcolor{black}{
    Therefore, the instantaneous reward is defined as the negative value with respect to the delay, i.e., 
    \begin{equation}{\label{equ:r}}
        \begin{aligned}
        &\mathcal{R}_{t+1}=\mathcal{R}_{i(t+1)}\!=\\
        &\!-\!10 \! \times\! \sum_{r \in R} \sum_{i,j \in \mathcal{U}}min(\mathcal{T}^{r}_{i(t) j(t)}, \mathcal{T}^{max}_{one}), t \in T,
    \end{aligned}
    \end{equation}
    where $\mathcal{T}^{r}_{i(t),j(t)}$ and $\mathcal{T}^{max}_{one}$ are shown in (\ref{delay}) and (\ref{delay_constraint}), respectively. }
    \item {\em Discount factor}: $\gamma_{i} \in (0,1)$ is the discount factor of agent $i$
    for calculating the cumulative reward. 
    $ \bm \gamma=\{\gamma_{i}, i \in \mathcal{U}\} $ is the discount factor set of all agents.
    A larger value of $\gamma$ indicates the decision focusing on the long-term reward.
    \end{itemize}

The policy $\pi_{i}$ leads agent $i$ to select an action 
$a_{i(t)}$ under observation $o_{i(t)}$.
$\bm \pi=\{\pi_{1},\pi_{2},\cdots ,\pi_{N}\}$ indicates the joint 
policy of all agents. 
According to the Dec-POMDP and a specific policy $\bm \pi$, 
we can obtain the routing paths. 
Hence, $\mathscr{P}0$ is transformed to find the optimal 
policy $\bm\pi^{\star}$, and the MADDQN-based intelligent approach is designed to 
obtain $\bm\pi^{\star}$ for minimizing the total end-to-end delay of routing paths.

\subsection{MADDQN Algorithm {\label{MADDQN}}}
The MADDQN is a Q-learning-based algorithm, which mainly 
constructs and maintains a Q-table containing the values of all state-action pairs. 
Formally, the Q-value function is defined as the expected 
total reward for taking action $a_{i(t)}$ in observation $o_{i(t)}$ for agent $i$ at time slot $t$, i.e.,
\begin{equation}
    Q(o_{i(t)},a_{i(t)})\!= \!\mathbb{E}\left[\sum_{k=0}^{\infty} \gamma_i^k \mathcal{R}_{i(t\!+k\!+1)} \! \mid \! o_{i(t)}\!=\!o, a_{i(t)}\!=\!a \right]\!,
\end{equation}
where $\mathbb{E}[\cdot]$ denotes the expected value. 
Through the Q-table, the optimal policy ${\pi^{\star}_i}$ can be obtained 
by agent $i$ in the process of continuous interactions with the environment.
In time slot $t$, the policy $\pi^{\star}_{i}$ is
\begin{equation}
    \pi^{\star}_{i}=\textrm{arg } \underset{{\pi_{i}}}{\textrm{max}}\;Q^{\star}(o_{i(t)},a_{i(t)}).
\end{equation}
Further, the following formula iteratively
updates the Q-value function $Q(o_{i(t)}, a_{i(t)}) $ via Bellman equations, i.e.,
\begin{multline}
        Q(o_{i(t)}, a_{i(t)}) \leftarrow (1-\alpha_{i})  Q(o_{i(t)}, a_{i(t)})  + \\
        \alpha_i \left[ \mathcal{R}_{i({t+1})} +  \gamma_i \max_{a^{\prime}_{i(t)}} Q(o_{i({t+1})}, a^{\prime}_{i(t)})\right], 
\end{multline}
where $\alpha_i$ is the learning rate that controls the 
size of update steps. $Q(o_{i({t+1})}, a^{\prime}_{i(t)})$ 
is the Q-value function of agent $i$ for taking action $a^{\prime}_{i(t)}$ 
under the next observation $o_{i(t+1)}$, 
which is transformed by performing action $a_{i(t)}$ in current observation $o_{i(t)}$
with reward $\mathcal{R}_{i(t+1)}$ in time slot $t$. 
The term $y=\mathcal{R}_{i(t+1)} + \gamma_i \max_{a^{\prime}_{i(t)}} Q(o_{i(t+1)}, a^{\prime}_{i(t)})$ 
represents the Q-target.

However, maintaining the Q-table becomes exceptionally complex,
when the dimensions of the state space and action space are large.
Hence, the DQN algorithm is proposed to approximate 
Q-value functions via deep neural networks (DNNs) with the observation as inputs and the 
action as outputs {\cite{Q-learning-based2}}. 
There exist two DNNs in DQN algorithms, i.e., 
the online network with parameter $\theta_{i}$ 
and the target network with parameter $\theta^{-}_{i}$ of each agent $i$.
By constantly updating the weight $\theta_i$ of online networks,
the loss function $L(\theta_{i})$ of agent $i$ is minimized  
and typically computed as the mean squared error between Q-value function
$Q(o_{i(t)}, a_{i(t)};  \theta_{i})$ and Q-target $y_{i(t)}^{\text{\tiny{DQN}}}$ 
via randomly sampling $B$ historical experience transitions from $\mathcal{B}_{i}$, i.e.,
\begin{equation}{\label{loss}}
    L(\theta_{i}) = \mathbb{E}_{\mathcal{B}_{i} \thicksim \varpi} \left[ \left( y_{i(t)}^{\text{\tiny{DQN}}}  
    - Q(o_{i(t)}, a_{i(t)};  \theta_{i}) \right)^2 \right],
\end{equation}  
and $y_{i(t)}^{\text{\tiny{DQN}}}$ is
\begin{equation}{\label{Q-target}}
    y_{i(t)}^{\text{\tiny{DQN}}} = \mathcal{R}_{i(t+1)}+ \gamma_{i} \max_{a^{\prime}_{i(t)}} Q(o_{i(t+1)} ,a^{\prime}_{i(t)};  \theta^{-}_{i}),
\end{equation}
where tuple $\varpi\! = \!(o_{i(t)}, a_{i(t)}, \mathcal{R}_{i(t+1)}, o_{i(t+1)})$ is a 
transition data stored in replay pool $\mathcal{B}_{i}$, 
which is designed to store the historical experience of agent $i$.
The set of experience replay pools for all agents is denoted as
 $\bm{\mathcal{B}}\!=\!\{\mathcal{B}_i, i \in \mathcal{U}\}$.

Nevertheless, the DQN-based algorithm may cause a large deviation,
due to overestimating the Q-target value. To avoid the
overestimation, a DDQN-based algorithm is leveraged by combining
double Q-learning with the DQN, and decouples the action selection 
and calculation of Q-target values. It is theoretically
proved that the DDQN-based approach can avoid the overestimation {\cite{van2016deep}}.
Correspondingly, Q-target $y_{i(t)}^{\text{\tiny{DQN}}}$ in loss 
function (\ref{loss}) is replaced by $y_{i(t)}^{\text{\tiny{DDQN}}}$,
which is
\begin{equation}{\label{DDQN-Q-target}}
    \begin{aligned}
    y_{i(t)}&^{\text{\tiny{DDQN}}}\! =\! \mathcal{R}_{i(t+1)}\! +\\
    &\! \gamma Q(o_{i(t+1)}, 
    \arg\max_{a^{\prime}_{i(t)}} Q(o_{i(t+1)}, a^{\prime}_{i(t)};  \theta_{i})|\theta^{-}_{i}).
  \end{aligned}
\end{equation}

\begin{algorithm}[t!]
    \caption{{\label{algorithm-MADDQN}}{MADDQN-based Routing Algorithm}}
    \begin{algorithmic}[1]
    \REQUIRE {$\mathcal{U}$, $R$, ${\bm{\alpha}}$, ${\bm{\gamma}}$, and $\mathcal{E}$}.
    \ENSURE Optimal policy ${\bm{\pi^{*}}}$.
    \STATE\textbf{Initialization:} {\label{initialize}}Initialize the UAV network environment, 
    hyper-parameters, and experience replay pool set $\bm {\mathcal{B}}$.
    \FOR{each episode} {\label{episode}}
    \FOR{$t=1, \dots, T$}
    \FOR{$i=1,\dots, N$}
    \STATE {\label{episode-initi}}The observation of agent $i$ is set as $o_{i(t)}$.
    \STATE {\label{o_a}}Agent $i$ selects an action $a_{i(t)}$ under observation $o_{i(t)}$ using an $\epsilon$-greedy policy.
    \STATE {\label{a_o}}Execute $a_{i(t)}$, and obtain reward $\mathcal{R}_{i(t+1)}$ and next observation $o_{i(t+1)}$.
    \STATE {\label{store}}Store the transition $(\!o_{i(t)}, \!a_{i(t)},\! \mathcal{R}_{i(t\!+\!1)},\! o_{i(t+1)}\!)$ 
    in the experience replay pool ${\mathcal{B}_i}$.
    \STATE {\label{update-s}} Set $ o_{i(t)}\leftarrow o_{i(t+1)}$.
    \IF{$|\mathcal{B}_{i}|>B$} {\label{sample}}
    \STATE Randomly select $B$ samples from the experience replay pool $\mathcal{B}_{i}$.
    \ENDIF
    \STATE {\label{target}} Compute the Q-target value $y_{i(t)}^{\text{\tiny{DDQN}}}$ for 
    each agent $i$ via ({\ref{DDQN-Q-target}}).
    \STATE {\label{update-theta}} Perform the gradient descent to update $\theta_{i}$ by ({\ref{theta-update}}). 
    \STATE {\label{update-theta-}} Periodically update the target network parameter $\theta^{-}_{i}$ via ({\ref{target-undate}}) every $\mathcal{W}$ steps.
    \ENDFOR
    \ENDFOR
    \ENDFOR
    \end{algorithmic}
\end{algorithm}

Besides, the gradient of loss functions is denoted 
as $\nabla_{\theta_{i}}L(\theta_{i})$,   
which is leveraged to  update $\theta_{i}$ 
via the gradient descent, i.e.,
\begin{equation}{\label{theta-update}}
    \theta_{i} \leftarrow \theta_{i} - \alpha_{i} \nabla_{\theta_{i}} L(\theta_{i}).
\end{equation}
Every $\mathcal{W}$ steps, parameter $\theta^{-}_{i}$ is periodically updated 
to match the parameter of online network $\theta_{i}$ for stabilizing training 
and improving the convergence, i.e.,
\begin{equation}{\label{target-undate}}
    \theta_{i}^{-}\leftarrow \theta_{i}.
\end{equation}

The detail of the proposed MADDQN is shown in Algorithm {\ref{algorithm-MADDQN}}.
Firstly, the algorithm initializes the UAV network environment,
hyper-parameters, and experience replay pool set $\bm{\mathcal{B}}$ ({line \ref{initialize}}).
At the beginning of each episode, the observation of each agent $i$ is initialized as $o_{i(t)}$ (lines {\ref{episode}}-{\ref{episode-initi}}).
Based on $o_{i(t)}$, each agent $i$ selects an action $a_{i(t)}$ by leveraging an 
$\epsilon$-greedy policy, and then executes $a_{i(t)}$ to obtain 
reward $\mathcal{R}_{i(t+1)}$ and next observation $o_{i(t+1)}$ (lines {\ref{o_a}}-{\ref{a_o}}). 
Hence, the agent $i$ obtains transition $(o_{i(t)},a_{i(t)}, \mathcal{R}_{i(t+1)}, o_{i(t+1)})$ and 
stores it into experience replay pool $\mathcal{B}_i$ (line {\ref{store}}).
Besides, the observation $o_{i(t)}$ of agent $i$ is correspondingly updated by $o_{i(t+1)}$ (line {\ref{update-s}}).
If the number of transition tuples in $\mathcal{B}_{i}$ is larger 
than the mini-batch $B$, $B$ samples are randomly selected to calculate 
Q-target value $y_{i(t)}^{\text{\tiny{DDQN}}}$ (lines {\ref{sample}}-{\ref{target}}). 
Additionally, the network performs a gradient descent step to update 
$\theta_{i}$ (line {\ref{update-theta}}). The target network parameter $\theta_{i}^{-}$ 
is periodically updated according to ({\ref{target-undate}}) every $\mathcal{W}$ steps (line {\ref{update-theta-}}).
    
\begin{table}[!t]
    \renewcommand\arraystretch{1.0}
    \begin{center}
       \caption{PARAMETER SETTING}
       {\label{table1}}
        \fontsize{9}{12}\selectfont{
        \begin{tabular}{|c|c||c|c|}
            \hline
            Parameter & Value & Parameter & Value \\ 
            \hline 
            $\vartheta$ & 2 & $f$ & 2.4GHz\\  
            \hline
            $c$ & $3 \times 10^{8}$ m/s & $\sigma_{ij}^{2}$ & -110dBm\\
            \hline 
            $B_{ij}$ & 2MHz & $M_{i}$ & 2kg \\
            \hline
            $g$ &  $\text{9.8m/s}^{2}$ & $\nu_{i}$ & 3m/s\\
            \hline 
            $\mathcal{P}_{b}^{hov}$ & 9.1827W & $\mathcal{P}_{I}^{hov}$ & 11.5274W \\
            \hline
            $U_{tip}$ & 60m/s  & $v_{{0}}$ & 2.4868m/s \\
            \hline
            $\partial _{0}$ & 0.5017  & $\jmath _{0}$ & 1.205$\text{kg/m}^3$ \\
            \hline
            $\imath _{0}$ & 0.0832  & $A_{0}$ & $\text{0.2827m}^2$ \\
            \hline
            $\beta  $ &  0.7 & $C^{max}_{i}$ & 50\\
            \hline
        \end{tabular}
        }
    \end{center}
\end{table}

\subsection{Complexity Analysis}
Suppose that the width of the $i$-th layer of the neural network is $W_i$, and there exist $M$ layers in total. 
$S$ and $A$ are the numbers of dimensions in the state space and the action space for an agent, respectively.
The computational complexity  from the \(i\)-th layer to the \((i + 1)\)-th 
layer involves the multiplication and summation of connection weights between neurons, 
which is \(O(W_i \cdot W_{i + 1})\). 
For the entire forward propagation process, the computational complexity of the agent is \(O(S\cdot A \cdot \sum_{i = 1}^{M - 1}W_i \cdot W_{i + 1})\), 
since the input and output layers are  $S$ and $A$, respectively.

In the routing process,
a single demand is transmitted from the source node to the destination node hop-by-hop. 
Assume that, on average, a demand requires \(h\) hops to reach the destination node. 
After \(h\) hops, the computational complexity of agents in the routing process is \( O(h\cdot S\cdot A\cdot\sum_{i = 1}^{M - 1}W_i \cdot W_{i + 1})\).
Since there exist \(n\) demands and the routing process of each demand is independent, 
the total computational complexity of the MADDQN algorithm is \(O(n \cdot h \cdot S\cdot A\cdot\sum_{i = 1}^{M - 1}W_i \cdot W_{i + 1})\).

\section{Simulation Results\label{sec:Simulation Results}}
In this section, we conduct a couple of simulations utilizing python. 
The specific parameters are listed in Table {\ref{table1}}
 {\cite{He_Routing,channel_model_PL,P_0_i}}. 
In particular, UAVs are distributed in the area within a 1.5km $\times$ 1.5km 
range, and the flight altitude range is within [0.12, 0.14]km. 
The minimum safe distance between UAVs is $d_{min}=$ 0.01km. 
The number of demands and the trust value of all UAVs are 
initialized randomly. Besides, the size of each demand is randomly 
set within [400, 600]kbits.
\textcolor{black}{$E_{\mathrm{elec}}$ and $\xi_{fs}$ are set as 
$1.5 \times 10^{-4}$ J/bit and  $2.5 \times 10^{-8}$  J/(bit$\times$m$^{2}$), respectively{\cite{commi_energy}}.}

\textcolor{black}{
To evaluate the designed adaptive weight methods, two baseline schemes are compared and detailed as follows.
\begin{itemize}
    \item \textit{Average weight methods}: The method indicates that $\psi _{i(t)}^1$ 
    and $\psi _{i(t)}^2$ are the average of $1-\psi _{i(t)}^0 $,
    while the calculation of $\psi _{i(t)}^0$ is the same as ({\ref{Adaptive_weights}}), i.e.,
    \begin{equation}
       \left\{  
        \begin{aligned}    
        \psi _{i(t)}^0  &= 0.5 \times \frac{\mathbb{T}_{thr}}{\mathbb{T}_{i(t)}}, \\
        \psi _{i(t)}^1 &= \psi _{i(t)}^2 = \frac{1-\psi _{i(t)}^0}{2}.
        \end{aligned}
        \right.
    \end{equation}
    \item \textit{Random weight methods}:
    The random weight method indicates that $\psi _{i(t)}^1$ is 
    a uniformly distributed random variable over the interval 
    $[0.2(1-\psi_{i(t)}^0), 0.8(1-\psi_{i(t)}^0)]$, i.e.,
    \begin{equation}
        \left\{
        \begin{aligned}
            \psi _{i(t)}^0 &= 0.5 \times \frac{\mathbb{T}_{thr}}{\mathbb{T}_{i(t)}}, \\
            \psi _{i(t)}^1 &\sim  {U}(0.2(1-\psi _{i(t)}^0), 0.8(1-\psi _{i(t)}^0)),  \\
            \psi _{i(t)}^2 &= 1-\psi _{i(t)}^0 - \psi _{i(t)}^1.
    \end{aligned} 
    \right.
    \end{equation}
\end{itemize}
Besides, to simulate the efficiency of the proposed 
adaptive weight method, the experimental configurations are as follows. 
\begin{itemize}
    \item The scale of UAV network is $N=20$, and the number of malicious nodes is $\mathtt{f}=2$.
    \item The initial trust value is set as 1, which represents that the UAVs are completely 
    trustworthy when departing from an initial station on the ground.
    Additionally, considering the high security requirements, the trust threshold is set as 0.8 {\cite{trust_parameters}}.
    \item According to {\cite{trust_parameters}}, the probability of malicious behaviors is set within $[0.1, 0.5]$. 
    Therefore, in the paper, the demand delivery probability and 
    correct path selection probability of malicious UAVs are characterized by 
    $p_1 \! \in \! \{0.5, 0.7, 0.9\}$ and $p_2 \! \in \! \{0.5, 0.7, 0.9\}$, respectively.
    Further, the simulation parameter space $\mathbb{P}$ of $(p_1,p_2)$ is
    \begin{equation}{\label{parameter-space}}
        \mathbb{P} = \begin{bmatrix}
            (0.5,0.5) & (0.5,0.7) & (0.5,0.9) \\
            (0.7,0.5) & (0.7,0.7) & (0.7,0.9) \\
            (0.9,0.5) & (0.9,0.7) & (0.9,0.9)
        \end{bmatrix}.
    \end{equation}
    \item The evaluation metric defined as the minimum time step that 
    can identify all malicious UAVs under various simulation parameter $(p_1,p_2)$ in $\mathbb{P}$, i.e.,
    \begin{equation}
        \Upsilon^{\text{detect}}_i = \min\{t \in \mathbb{N}^+ | \mathbb{T}_{i(t)}<\mathbb{T}_{thr}\}, \forall i \in \mathcal{U}, t\in T.
    \end{equation}
\end{itemize}
}

\begin{figure}[t]
    \centering
    \includegraphics[width=0.8\linewidth]{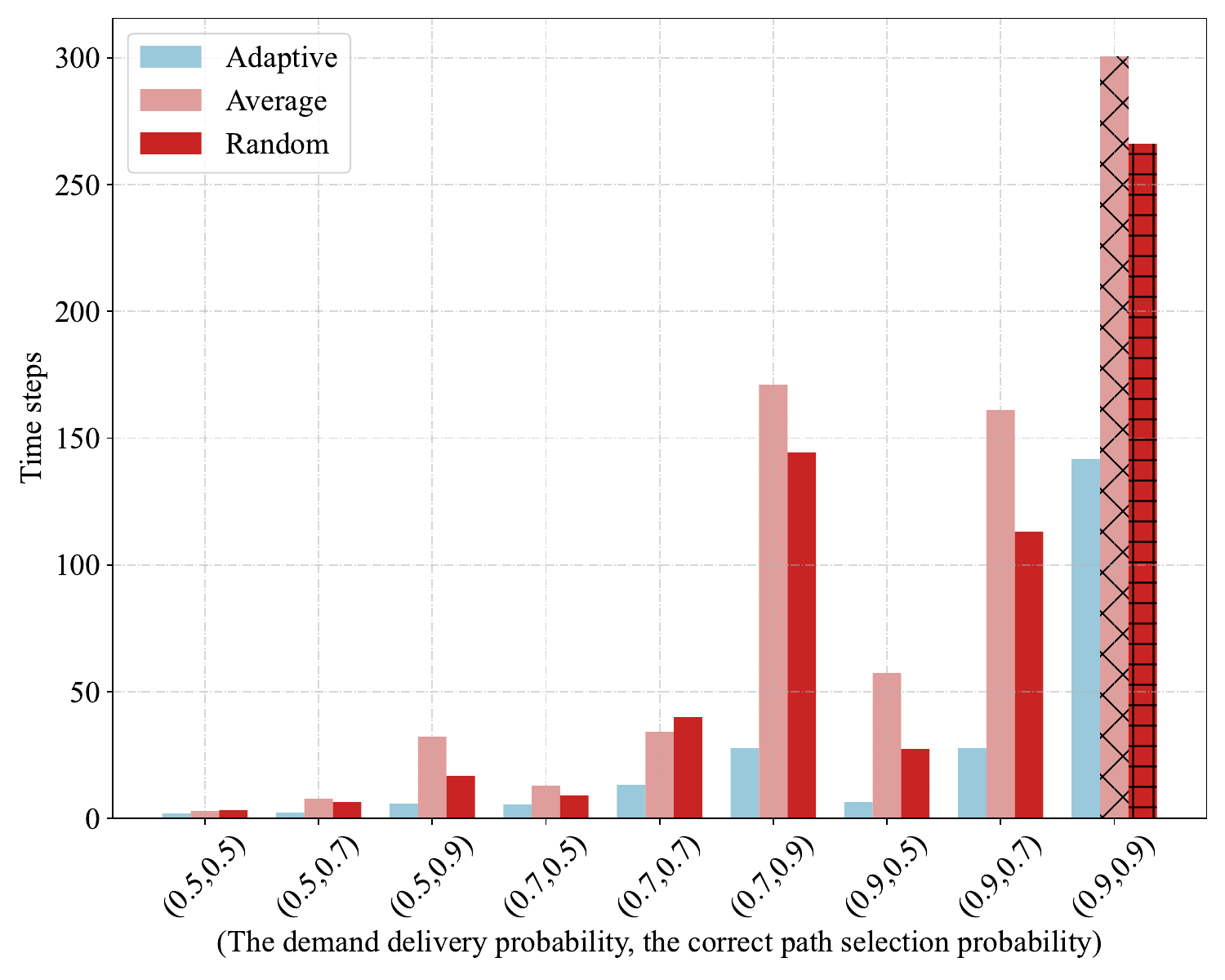}
    \vspace{-0.3cm}
    \textcolor{black}{\caption{ The minimum time step for identifying malicious UAVs with different simulation parameters in space $\mathbb{P}$.}
  \label{fig:trust-value}}
  \vspace{-0.3cm}
 \end{figure}

 \begin{figure*}[t]
    \vspace{-0.3cm}
    \centering
    \subfloat[\textcolor{black}{ }]{\centering
    \includegraphics[width=0.33\linewidth]{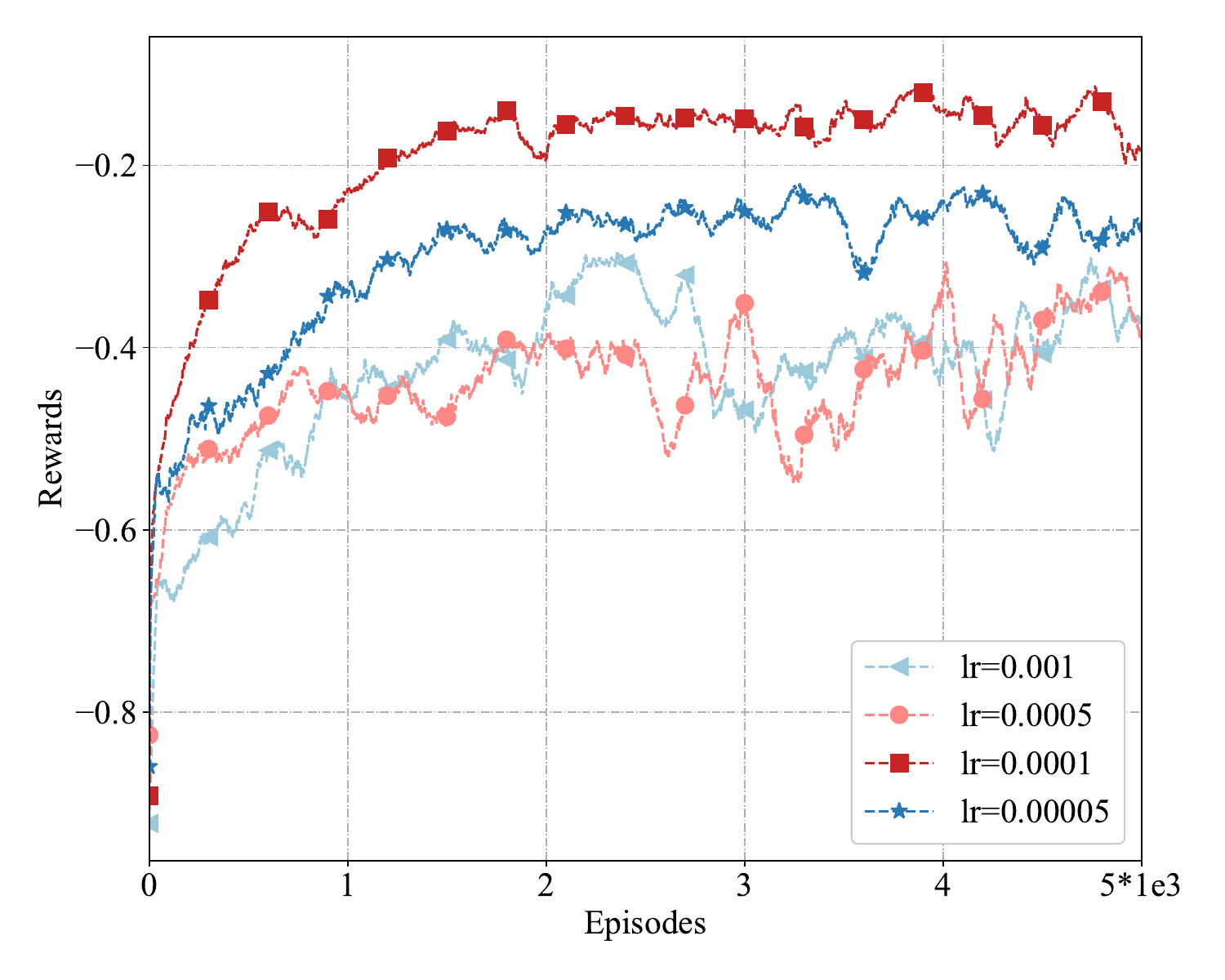}}
    \subfloat[\textcolor{black}{ }]{\centering
    \includegraphics[width=0.33\linewidth]{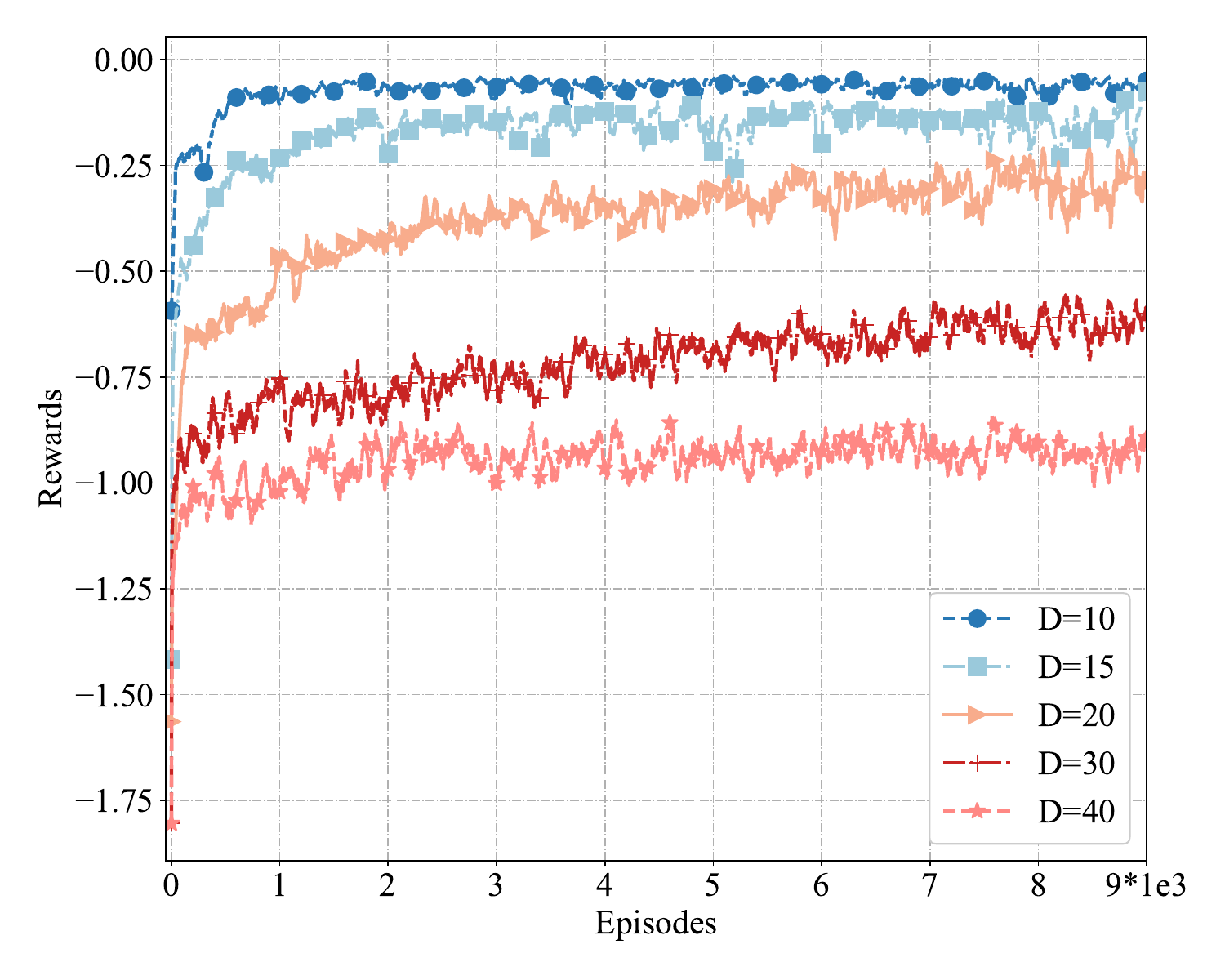}}
    \subfloat[\textcolor{black}{}]{\centering
    \includegraphics[width=0.33\linewidth]{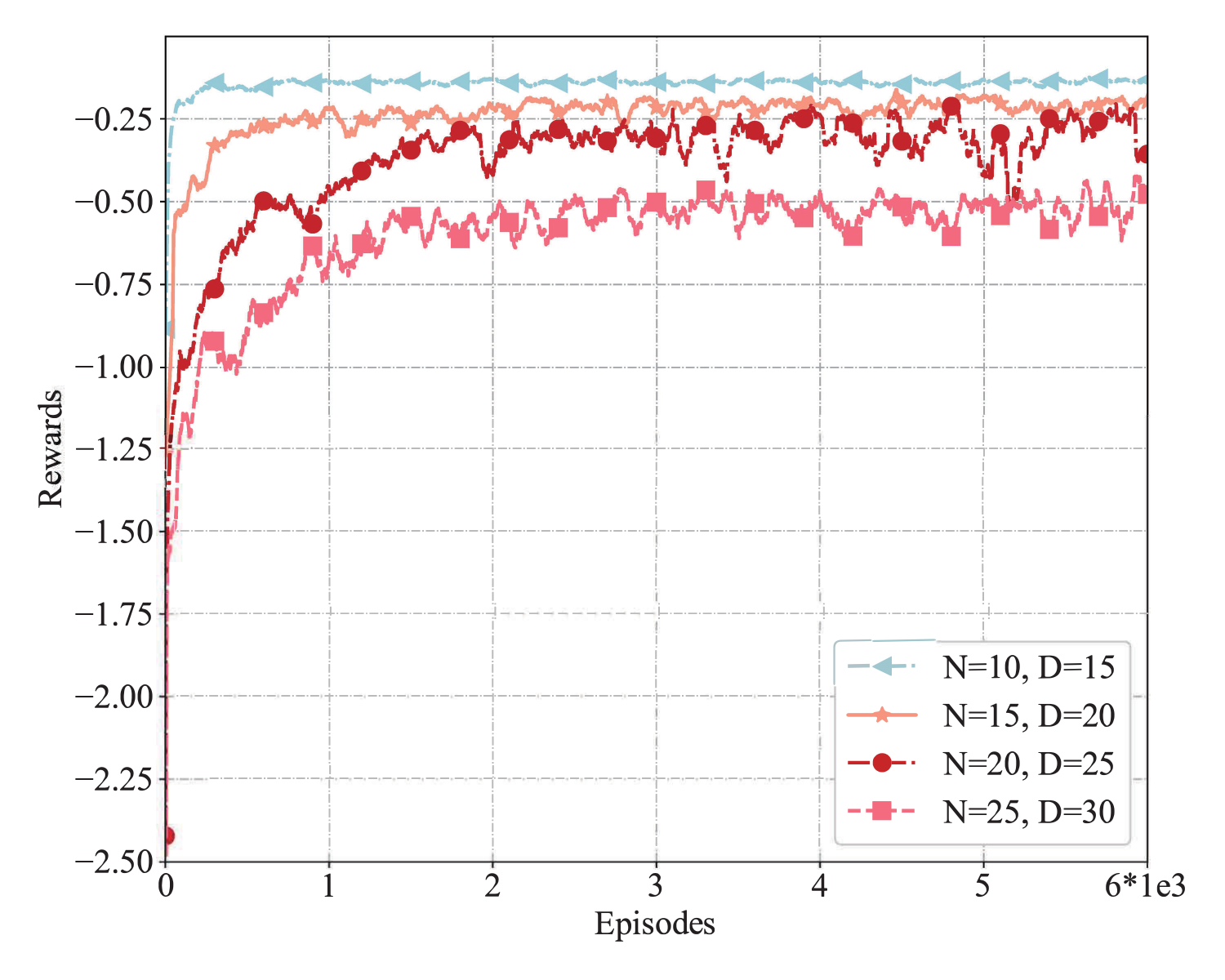}}
    \caption{Comparisons of the accumulative reward of episodes obtained by total UAVs for transmitting 
    the given demands during the training process with 2 malicious UAVs.
    (a) Different learning rates with the same 20 UAVs and 25 demands. 
    (b) Different demand numbers with the same 10 UAVs.
    (c) Different UAV numbers with different demand numbers.}
    {\label{fig:lr-number-demand}}
    \vspace{-0.4cm}
\end{figure*}

\begin{figure*}[t]
    \centering
    \subfloat[\textcolor{black}{ }]{\centering
    \includegraphics[width=0.325\linewidth]{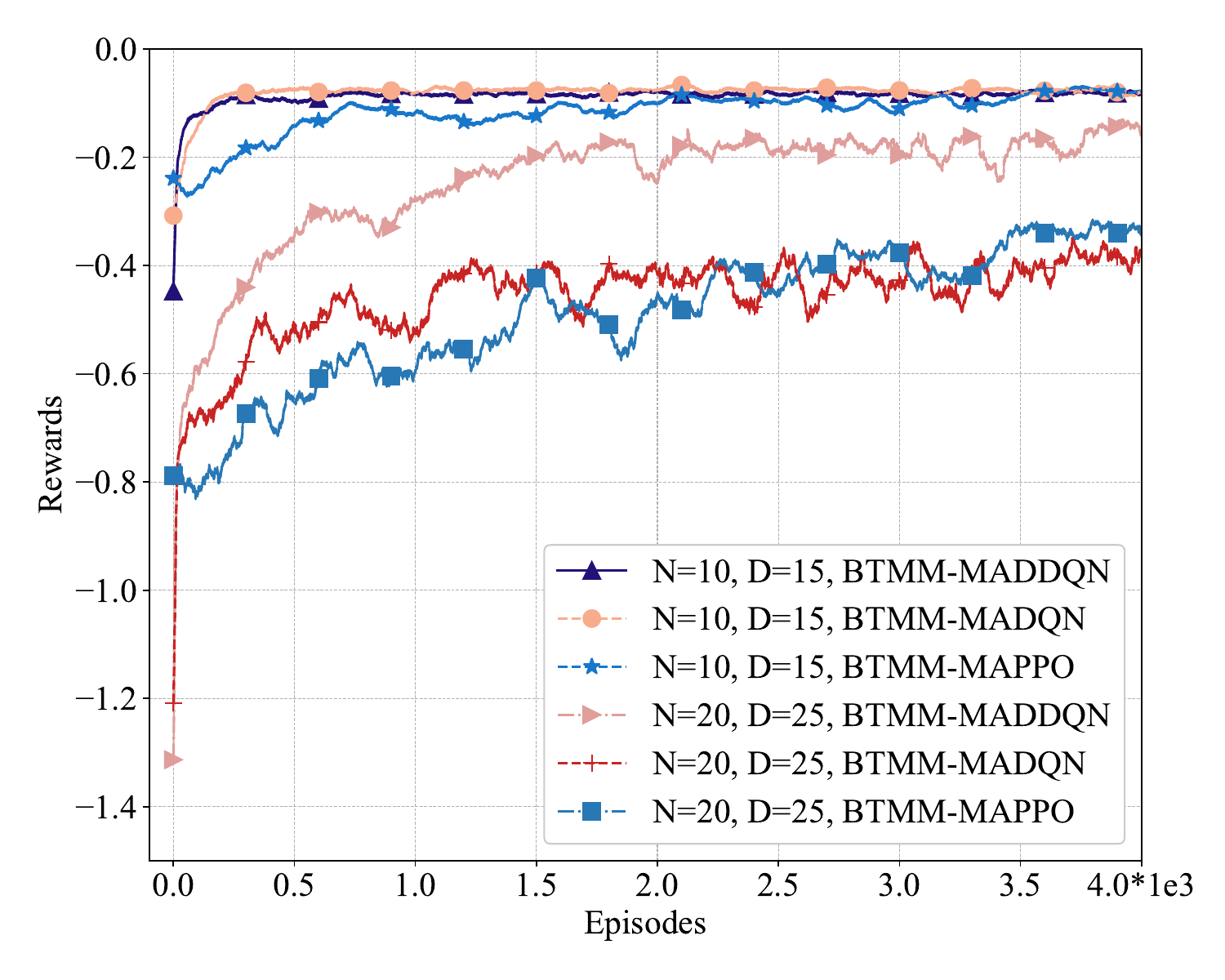}}
    \subfloat[\textcolor{black}{ }]{\centering
    \includegraphics[width=0.325\linewidth]{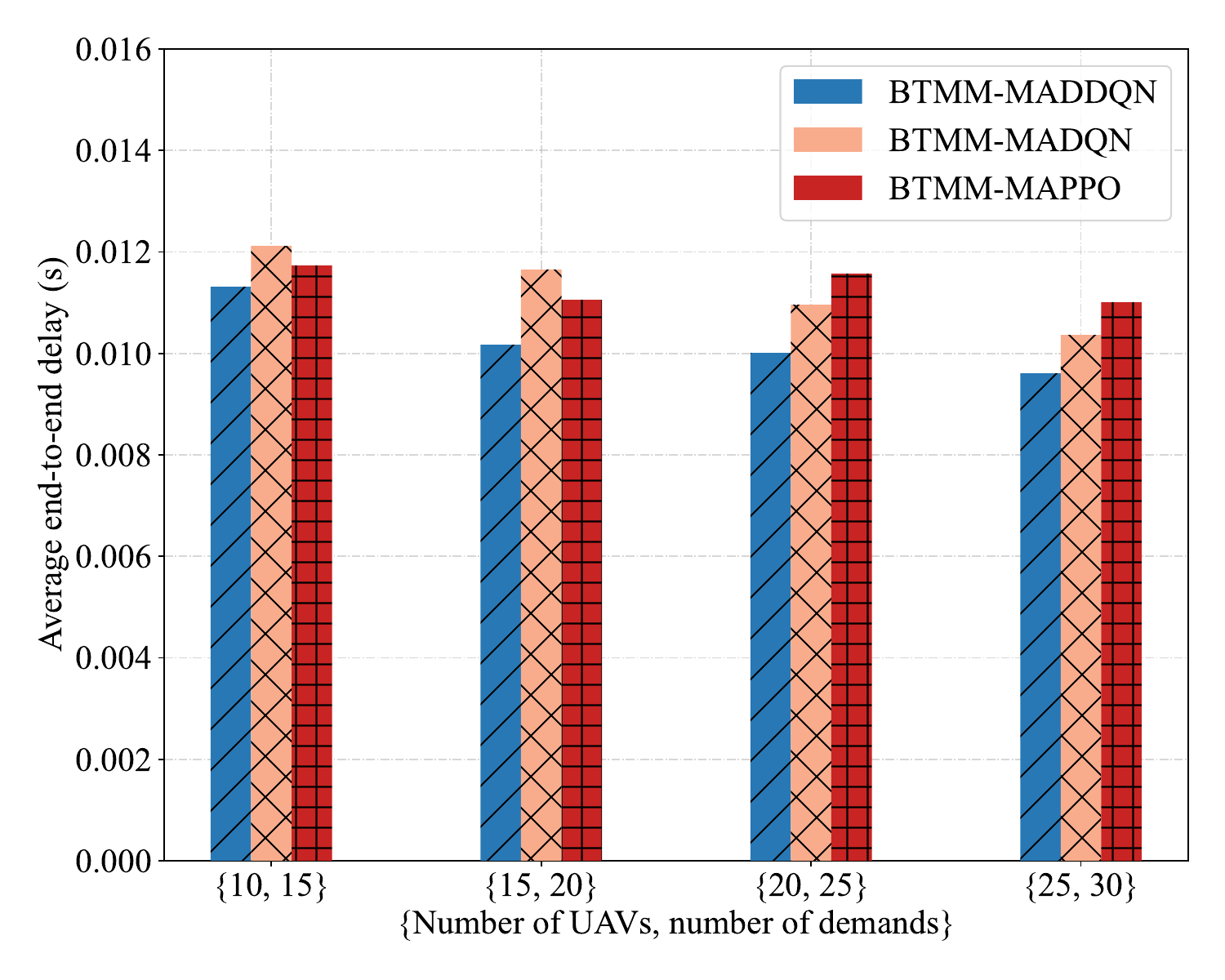}}
    \subfloat[\textcolor{black}{ }]{\centering
    \includegraphics[width=0.322\linewidth]{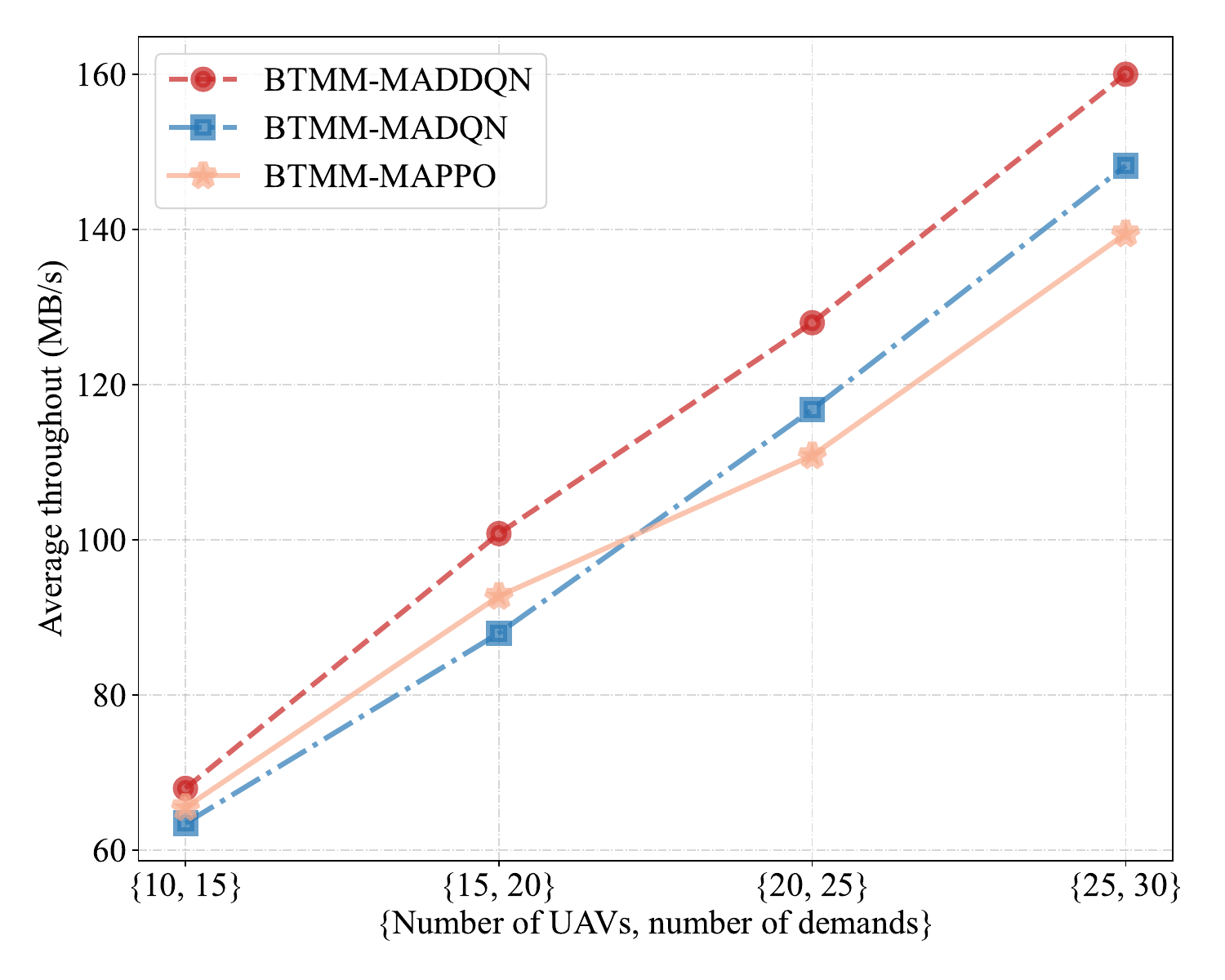}}
    \caption{Comparisons of the three algorithms in terms of the convergence performance,
     delay cost, and throughput with 2 malicious UAVs, respectively.
     (a) Rewards of episodes with different scales of UAV networks.
     (b) Average end-to-end delay with different UAV and demand numbers.
     (c) Average throughput with different UAV and demand numbers.
     }
    {\label{fig:3Algorithms}}
    \vspace{-0.3cm}
\end{figure*}
\textcolor{black}{
In Fig. {\ref{fig:trust-value}}, the proposed adaptive weight method calculating 
the trust value is compared with the average and random weight methods. 
Specifically, we assess the metric of the minimum time steps required to identify 
all malicious UAVs, across a diverse range of simulation parameters in (\ref{parameter-space}).  
It can be observed that when $(p_1,p_2)\!=\!(0.5, 0.5)$, it needs the fewest time steps for identifying the malicious UAVs. 
In contrast, when $(p_1,p_2)\!=\!(0.9, 0.9)$, the most time steps are required. 
The reason lies that the decreasing values of $p_1$ and $p_2$ indicate that the incorrect behaviors of malicious UAVs occur more frequently, 
leading to the increment of the evaluated incorrect behavior rates 
 (i.e., $1-\mathbb{T}^{dr}_{i(t)}$ and $1-\mathbb{T}^{tp}_{i(t)}$ in formula (\ref{Adaptive_weights})). 
 Besides, it is noted that the adaptive method consistently obtains the minimum time steps compared with 
 the other methods in various simulation parameters.
 This is due to the fact that the trust weights in the proposed method are proportional to incorrect behavior rates, 
 and thus the trust values of UAVs are dynamically adjusted at a quadratic rate. 
 The quadratic variation can impose more severe penalties on incorrect behaviors and accelerate the detection of malicious UAVs.
 Therefore, under the same case, the proposed adaptive weight method performs better than both the average and random weight methods.
 }

In Fig. {\ref{fig:lr-number-demand}}, we provide  
accumulative rewards to evaluate the convergence performance of the proposed 
BTMM-MADDQN algorithm under different learning rates, UAV numbers, 
and demand numbers, respectively. Besides, the number of malicious UAVs is 2.
In Fig. {\ref{fig:lr-number-demand}}(a), for all learning rates, the number of UAVs is 20 
and the number of demands is set as 25. 
It can be observed that different learning rates have different 
performances on the convergence. In the first 2,000 episodes, 
rewards are not satisfactory. As the number of episodes increases, 
rewards increase and converge at the specific values. 
In detail, the reward at the learning rate of 0.0001 outperforms 
the rewards of other learning rates. 
Besides, when the value is 0.0005, the convergence performance is the worst 
and does not converge to a relatively stable result, which illustrates that a large 
learning rate may lead to a local solution. Considering the execution effect 
analyzed above, a learning rate of 0.0001 is selected  
for the rest of simulations.

In the network with 20 UAVs, Fig. {\ref{fig:lr-number-demand}}(b)
shows the rewards of the BTMM-MADDQN algorithm with different demand numbers.
It is noted that the case of 40 demands obtains 
the slowest convergence rate and the smallest reward.
Meanwhile, as the number of demands decreases, the convergence rate is faster and the reward is larger.
Since when there exist more demands, 
the environment perceived by UAVs is more complex and they require more episodes to learn,
leading to a slower convergence speed. Besides, as the reward designed in ({\ref{equ:r}})
is related to the negative value of the summed delay for transmitting all demands, 
the more demands obtain the lower rewards.

In Fig. {\ref{fig:lr-number-demand}}(c), the reward is evaluated in different scales of UAV networks
with different demand numbers.
Although the increased number of UAVs provides more resources to serve, the reward value decreases as the number of demands 
increases, since the reward is the sum of delay generated by all UAVs transmitting all demands.  
Besides, the convergence is slower with the scale of UAV networks increases and the demand number grows, 
since each agent needs more episodes to learn the more complex information of neighbor UAVs.
In short, Fig. {\ref{fig:lr-number-demand}} illustrates 
that the BTMM-MADDQN algorithm effectively finds routing paths via maximizing 
convergent reward values, under various learning rates, UAV network scales, and demand numbers.

In Fig. {\ref{fig:3Algorithms}}, the BTMM-MADDQN is compared with BTMM-MAPPO and 
BTMM-MADQN algorithms with regard to the convergence performance, 
average end-to-end delay, and average throughput.
Specifically, in Fig. {\ref{fig:3Algorithms}}(a), the rewards of all three algorithms converge. 
When the numbers of UAVs and demands are same, for BTMM-MADDQN, the convergence speed is faster 
and rewards are larger compared to the BTMM-MAPPO and BTMM-MADQN algorithms. 
Specially, in networks with 10 UAVs or 20 UAVs, the effects of BTMM-MAPPO are the worst 
among the three algorithms.
The end-to-end delay for transmitting demands is 
evaluated in Fig. {\ref{fig:3Algorithms}}(b). 
It is observed that as the network becomes larger and demands grow, 
the delay of data transmission increases in all three algorithms. 
Nevertheless, the delay of the proposed BTMM-MADDQN algorithm decreases by
13.39$\%$ and 12.74$\%$ than the BTMM-MAPPO algorithm at $\{20,25\}$ 
and the BTMM-MADQN algorithm at $\{15,20\}$, respectively.
Fig. {\ref{fig:3Algorithms}}(c) shows the performance of throughput 
in networks with different UAV and demand numbers. 
As the number of UAVs increases, the communication channel resources of networks are richer, 
and thus more demands are transmitted per second, leading to an increment in throughput.
When the UAV number is 10, the network has the lowest throughput of 67.94MB/s for the proposed algorithm.
Besides, the throughput of the BTMM-MADDQN algorithm increases by 
15.44$\%$ and 14.6$\%$ than the BTMM-MAPPO algorithms at $\{20,25\}$ and BTMM-MADQN 
algorithms at $\{15,20\}$, respectively.

\begin{figure}[t]
    \centering
    \includegraphics[width=0.8\linewidth]{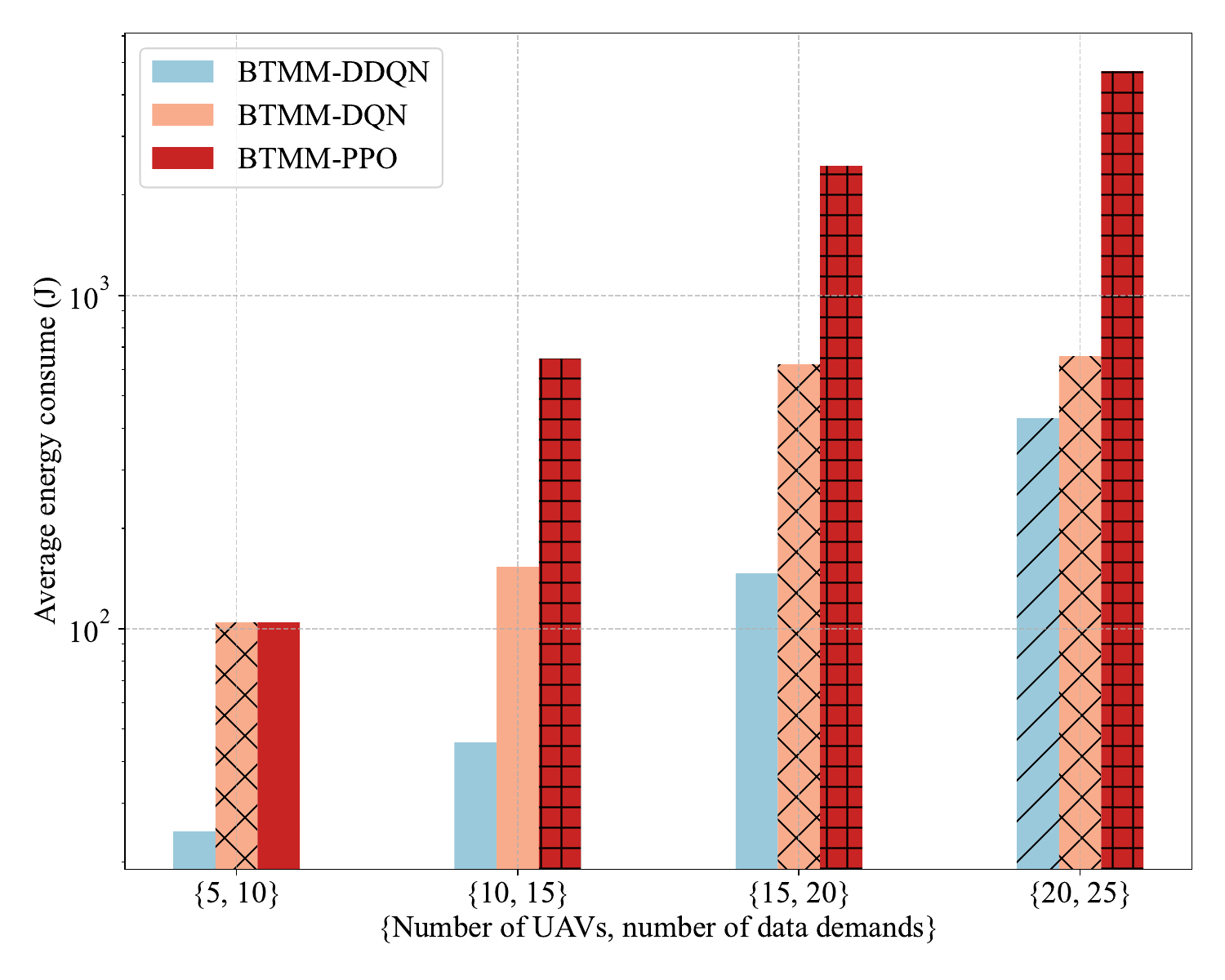}
    \vspace{-0.3cm}
    \textcolor{black}{\caption
    {\label{fig:energy-consume}The energy consumption of UAVs with different UAV numbers and demand numbers under
    different algorithms.}
    }
    \vspace{-0.4cm}
\end{figure}

\textcolor{black}{
    Fig. {\ref{fig:energy-consume}} illustrates the average energy consumption of UAVs with different scales of networks and demands.
    It is noted that as the number of demands increases, it requires more resources to complete the requirement, 
    leading to the increased energy consumption. 
    However, the energy consumption of the proposed BTMM-MADDQN is less 
    than other algorithms, due to the less delay required by the demands. 
    In detail, when the transmission delay is shorter, demands can reach the destination UAVs with the greater speed and precision,
    reducing the energy consumption caused by possible suboptimal routing. Besides, it is beneficial for 
    each UAV to obtain the information of other UAVs timely, strengthening the collaborative operation among UAVs, 
    and thus avoiding the unnecessary energy waste.
}

\begin{figure}[t]
    \centering
    \includegraphics[width=0.8\linewidth]{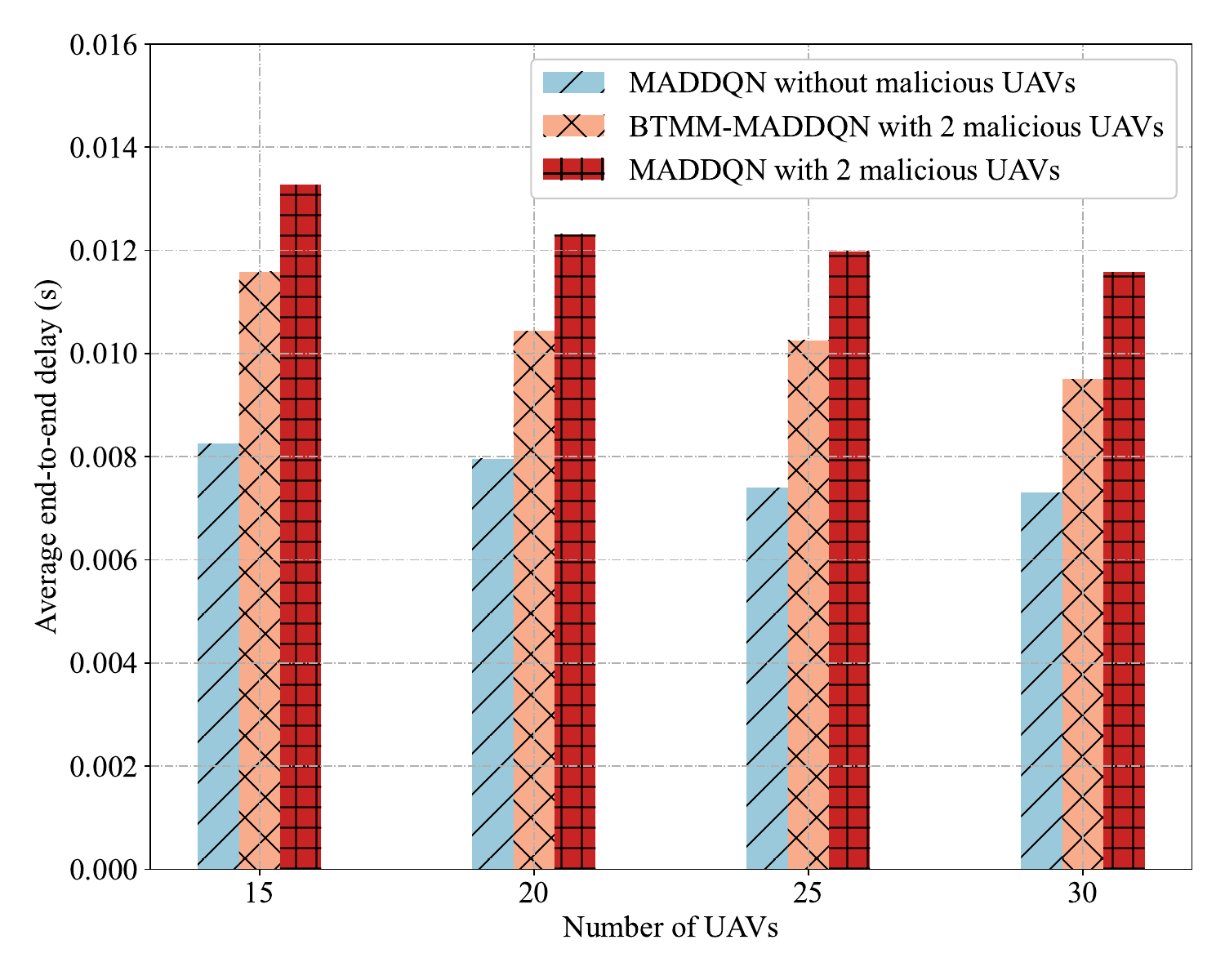}
    \vspace{-0.3cm}
    \caption{Average end-to-end delay with different UAV numbers, under 
    different algorithms and malicious UAV numbers.
  \label{fig:Deliberate-Delay1}}
  \vspace{-0.4cm}
\end{figure}

With different UAV network scales, Fig. {\ref{fig:Deliberate-Delay1}} verifies the superiority of 
the proposed BTMM-MADDQN algorithm in terms of delay when there exist malicious UAVs. Specifically,
in the UAV network without malicious nodes, the MADDQN-based routing 
method shows the excellent low delay performance, since UAVs do not generate
malicious behaviors such as transmitting data via error paths with extra delay.
However, in UAV networks with malicious nodes, the MADDQN-based approach 
obtains the worst performance of delay, since malicious UAVs may transmit 
demands under non-optimum paths, leading to the increased delay cost.
Meanwhile, we can observe that the delay of the BTMM-MADDQN-based algorithm
decreases by 16.6$\%$ than MADDQN-based approaches when there exist malicious UAVs.

\begin{figure}[t]
    \centering
    \includegraphics[width=0.8\linewidth]{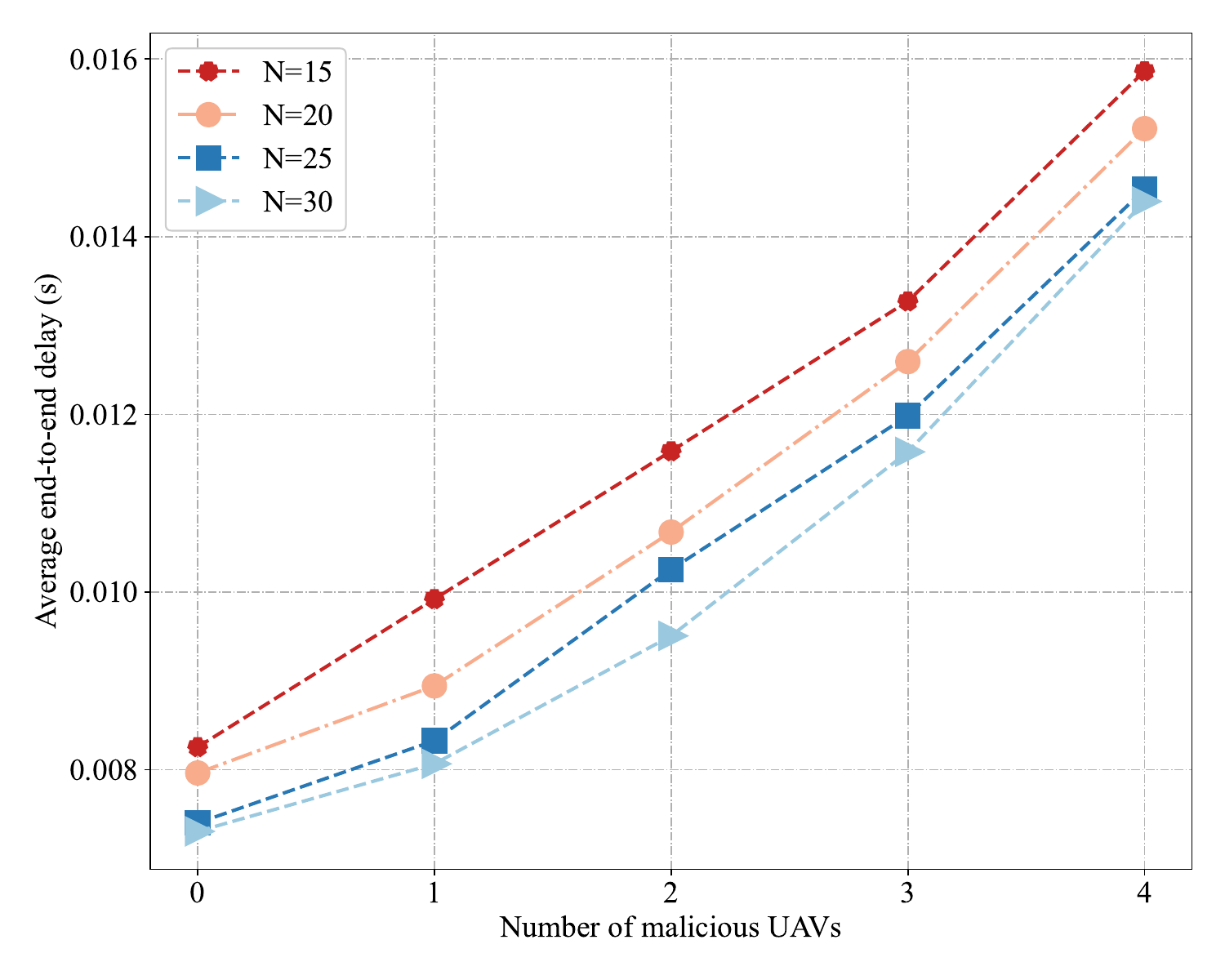}
    \vspace{-0.3cm}
    \caption{Average end-to-end delay of the proposed BTMM-MADDQN with different 
    network scales and malicious UAV numbers. 
  \label{fig:Deliberate-Delay2}}
  \vspace{-0.2cm}
\end{figure}
\begin{figure}[t]
    \centering
    \subfloat[\textcolor{black}{ }]{\centering
    \includegraphics[width=0.8\linewidth]{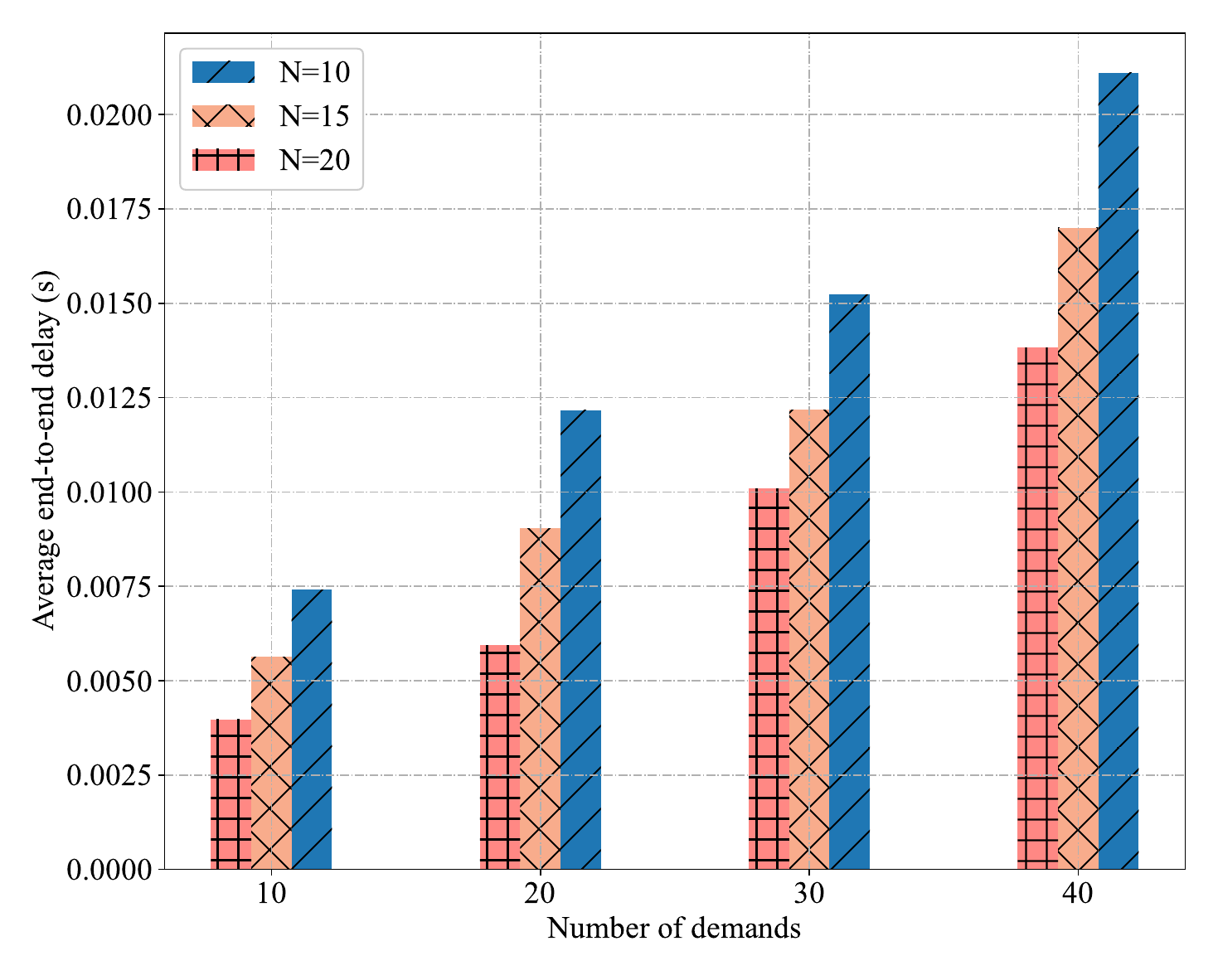}}
    \vspace{-0.4cm}
    \vfill
    \subfloat[\textcolor{black}{ }]{\centering
    \includegraphics[width=0.8\linewidth]{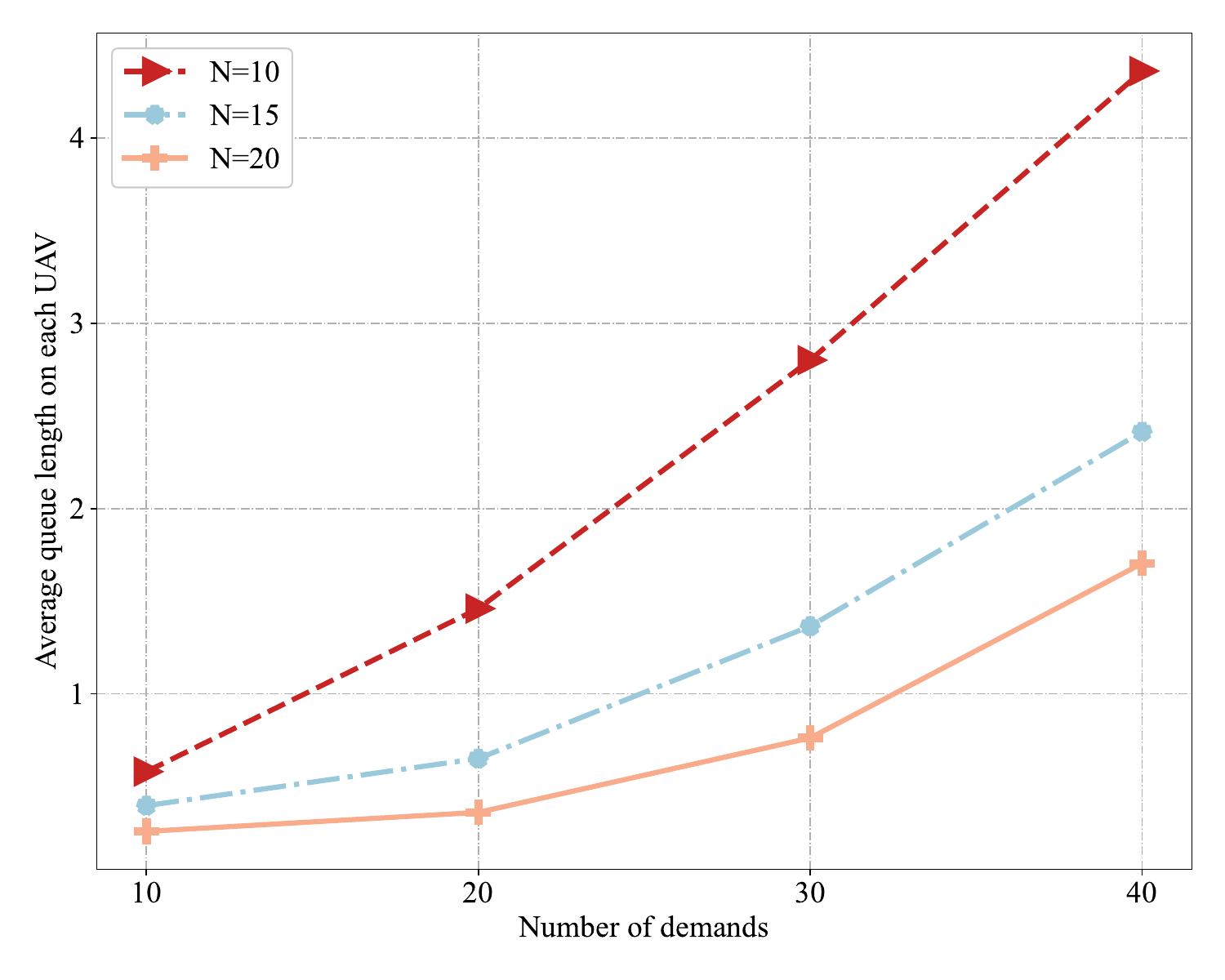}}
    \caption{Comparisons of different scales of UAV networks with 2 malicious UAVs.
    (a) Average end-to-end delay cost with different demand numbers.
    (b) The average queue length on each UAVs with different demand numbers.
  \label{fig:Demand-Number}}
  \vspace{-0.3cm}
\end{figure}

Fig. {\ref{fig:Deliberate-Delay2}} illustrates the performance of delay 
when there exist different numbers of malicious UAVs in different UAV 
network scales. It can be seen that the delay is the lowest in networks without malicious UAVs, 
while there exist malicious UAVs, the delay increases significantly.
It indicates that malicious UAVs in routing paths reduce the delay performance.
Besides, as the number of  malicious UAVs increases, the delay grows sharply. 
It is accounted by the fact that the original routing paths 
are corrupted and replaced by suboptimal routing paths. 
Meanwhile, as the number of UAVs becomes larger, 
the total network resources are richer in terms of the energy, 
channel, and queue capacity, resulting in the lower delay cost of routing.

In Fig. {\ref{fig:Demand-Number}}, the impacts of demands and UAVs on the routing 
performance are evaluated.
Specifically, Fig. {\ref{fig:Demand-Number}}(a) depicts the delay cost of different 
demand numbers with different numbers of UAVs. As the demand number increases, the total end-to-end delay 
rises, due to the increased congestion in UAV networks. 
Nevertheless, with the same demands, the delay decreases 
as the UAV number increases, since the network service is more resourced and 
the demand  is transmitted at a high probability without queuing on UAVs. 
Fig. {\ref{fig:Demand-Number}}(b) depicts the average queue length 
on UAVs for different demands in networks. 
It is noted that as the number of demands increases, the queue length rises quickly,
due to the increased congestion in UAV networks. 
Meanwhile, for the same amount of the demand to be transmitted in networks, 
the average queue length is smaller when UAV numbers are larger, 
due to the more adequate service resources.
Besides, when the demand numbers are greater than UAV numbers, 
the average queue length on each UAV increases at a greater speed.

\section{Conclusions\label{sec:Conclusions}}
In this work, we depict the routing process in 
a time-varying UAV network with malicious nodes. Besides, the routing problem is 
formulated to minimize the total end-to-end delay with multi-constraints. 
To enhance the safety during routing processes, we put forward the BTMM, 
in which the node trust evaluation mechanism is designed, 
and consensus UAV update mechanisms are proposed to improve the security of
PBFT algorithms. 
Further, to deal with the challenge of obtaining global information
in decentralized UAV networks, we reformulate the routing problem into 
a Dec-POMDP, and an MADDQN-based algorithm is proposed to solve it. 
Simulation results in UAV networks with attacked nodes demonstrate that 
the designed BTMM-MADDQN algorithm outperforms
in multi aspects, such as the convergence performance, delay, throughput, 
and queue length, and in particular the proposed algorithm  decreases 
delay by 16.6$\%$ than methods without the BTMM.
\textcolor{black}
 {
     \bibliographystyle{IEEEtran}
     \bibliography{ref2}
 }
\end{document}